\documentclass[traditabstract, twocolumns, longauth]{aa}  

\usepackage{txfonts}
\usepackage{graphicx}	% Including figure files
\usepackage{amsmath}	% Advanced maths commands
\usepackage{amssymb}	% Extra maths symbols
\usepackage{xcolor}     % Text color
\usepackage{soul}       % for strikeout, can be removed after comments are treated
\usepackage{array}
\def\ksmpc{${\, \mathrm{km}\, \mathrm{s}^{-1}\, \mathrm{Mpc}^{-1}}$\xspace}
\def\ks{${\, \mathrm{km}\, \mathrm{s}^{-1}}$\xspace}
\newcommand{\lcdm}{$\mathrm{\Lambda CDM}$\xspace}
\newcommand{\slope}{$\xi$}
\newcommand{\Bsixteen}{B1608$+$656\xspace}
\newcommand{\RXJ}{RX\ J1131$-$1231\xspace}
\newcommand{\HEzero}{HE\ 0435$-$1223\xspace}
\newcommand{\Jtwelve}{SDSS\ J1206$+$4332\xspace}
\newcommand{\WFItwenty}{WFI\ 2033$-$4723\xspace}
\newcommand{\PGeleven}{PG\ 1115$+$080\xspace}
\newcommand{\DESzerofour}{DES\ J0408$-$5354\xspace}
\newcommand{\LENSTRO}{{\tt LENSTRONOMY}}

\newcommand{\Hc}{\ensuremath{H_0}\xspace}
\newcommand{\zd}{\ensuremath{z_\mathrm{d}}\xspace}

\newcommand{\Dd}{\ensuremath{D_\mathrm{d}}\xspace}
\newcommand{\Dds}{\ensuremath{D_\mathrm{ds}}\xspace}
\newcommand{\Ds}{\ensuremath{D_\mathrm{s}}\xspace}
\newcommand{\Ddt}{\ensuremath{D_{\Delta t}}\xspace}
\newcommand{\kext}{\ensuremath{\kappa_\mathrm{ext}}\xspace}
\newcommand{\raper}{\ensuremath{\theta_\mathrm{aperture}}\xspace}
\newcommand{\reff}{\ensuremath{\theta_\mathrm{eff}}\xspace}
\newcommand{\thetaE}{\ensuremath{\theta_\mathrm{E}}\xspace}

\begin{document}

\title{
TDCOSMO. I. An exploration of systematic uncertainties in the inference of \Hc from time-delay cosmography 
%TDCOSMO. I. An exploration of systematic uncertainties in time-delay cosmography and the inference of \Hc
%Asserting the impact of galaxy kinematics on \Hc measurements with time-delays in gravitationally lensed quasars
}
\author{
%group 1
M.~Millon\inst{\ref{epfl}} \and
A.~Galan\inst{\ref{epfl}} \and
F.~Courbin\inst{\ref{epfl}} \and
T.~Treu\inst{\ref{ucla}} \and
S.~H.~Suyu\inst{\ref{mpa},\ref{tum},\ref{asiaa}} \and
X.~Ding\inst{\ref{ucla}} \and
%group 2 (alphabetical)
S.~Birrer\inst{\ref{stanford}} \and
G.~C.-F.~Chen \inst{\ref{ucdavis}} \and
A.~J.~Shajib \inst{\ref{ucla}} \and
D.~Sluse\inst{\ref{star}} \and
K.~C.~Wong \inst{\ref{kavliipmu}} \and
% group 3 (alphabetical)
A.~Agnello\inst{\ref{dark}} \and
M.~W.~Auger\inst{\ref{ucambridge},\ref{kavlicambridge}} \and
E.~J.~Buckley-Geer\inst{\ref{Fermilab}} \and
J.~H.~H.~Chan\inst{\ref{epfl}}\and
T.~Collett\inst{\ref{uportsmouth}} \and
C.~D.~Fassnacht \inst{\ref{ucdavis}} \and
S.~Hilbert \inst{\ref{mpa}} \and
L.~V.~E.~Koopmans\inst{\ref{unigroningen}}\and 
V.~Motta \inst{\ref{universidadvalparaiso}} \and
S.~Mukherjee \inst{\ref{star}}\and
C.~E.~Rusu \inst{\ref{naoj}} \and
A.~Sonnenfeld \inst{\ref{leiden}} \and
C.~Spiniello \inst{\ref{OAC},\ref{ESO}} \and
L.~Van de Vyvere\inst{\ref{star}} 
}

\institute{
Institute of Physics, Laboratory of Astrophysics, Ecole Polytechnique 
F\'ed\'erale de Lausanne (EPFL), Observatoire de Sauverny, 1290 Versoix, 
Switzerland \label{epfl} \goodbreak 
\and
Department of Physics and Astronomy, University of California, Los Angeles CA 90095, USA \label{ucla} \goodbreak
\and
Max-Planck-Institut f{\"u}r Astrophysik, Karl-Schwarzschild-Str.~1, 85748 Garching, Germany \label{mpa} \goodbreak
\and
Physik-Department, Technische Universit\"at M\"unchen, James-Franck-Stra\ss{}e~1, 85748 Garching, Germany \label{tum} \goodbreak
\and
Academia Sinica Institute of Astronomy and Astrophysics (ASIAA), 11F of ASMAB, No.1, Section 4, Roosevelt Road, Taipei 10617, Taiwan \label{asiaa} \goodbreak
\and 
Kavli Institute for Particle Astrophysics and Cosmology and Department of Physics, Stanford University, Stanford, CA 94305, USA \label{stanford}\goodbreak
\and
 Department of Physics, University of California, Davis, CA 95616, USA \label{ucdavis}\goodbreak
\and 
STAR Institute, Quartier Agora - All\'ee du six Ao\^ut, 19c B-4000 Li\`ege, Belgium \label{star} \goodbreak
 \and
 Kavli IPMU (WPI), UTIAS, The University of Tokyo, Kashiwa, Chiba 277-8583, Japan\label{kavliipmu} \goodbreak
 \and
DARK, Niels Bohr Institute, Lyngbyvej 2, 2100 Copenhagen, Denmark \label{dark} \goodbreak
\and 
 Kapteyn Astronomical Institute, University of Groningen, P.O.Box 800, 9700AV Groningen, Netherlands \label{unigroningen} \goodbreak
 \and 
 Instituto de F\'{\i}sica y Astronom\'{\i}a, Facultad de Ciencias, Universidad de Valpara\'{\i}so, Avda. Gran Breta\~na 1111, Valpara\'{\i}so, Chile. \label{universidadvalparaiso} \goodbreak
 \and
 National Astronomical Observatory of Japan, 2-21-1 Osawa, Mitaka, Tokyo 181-8588, Japan \label{naoj}\goodbreak
\and
Institute of Astronomy, University of Cambridge, Madingley Road, Cambridge CB3 0HA, UK \label{ucambridge}\goodbreak
\and 
Kavli Institute for Cosmology, University of Cambridge, Madingley Road, Cambridge CB3 0HA, UK \label{kavlicambridge}\goodbreak
\and
Leiden Observatory, Leiden University, Niels Bohrweg 2, 2333 CA Leiden, the Netherlands
\label{leiden}\goodbreak
\and
University of Portsmouth, Institute of Cosmology and Gravitation, Portsmouth PO1 3FX, United Kingdom \label{uportsmouth} \goodbreak
\and
INAF - Osservatorio Astronomico di Capodimonte, Salita Moiariello, 16, 80131, Napoli, Italy
\label{OAC} \goodbreak
\and
European Southern Observatory, Karl-Schwarschild-Str. 2, 85748, Garching, Germany
\label{ESO} \goodbreak
\and 
Fermi National Accelerator Laboratory, P. O. Box 500, Batavia, IL 60510, USA \label{Fermilab}
}

%\datesmooth{\today}
\abstract{
Time-delay cosmography
%SHS added:
of lensed quasars
has achieved
% Perhaps add "better than" (CDF)
2.4\% precision on the measurement of the Hubble constant, \Hc. As part of an ongoing effort to uncover and control systematic uncertainties, we investigate three potential sources: 1- stellar kinematics, 2- line-of-sight effects, and 3- the deflector mass model. To meet this goal in a quantitative way, we reproduced the H0LiCOW/SHARP/STRIDES (hereafter TDCOSMO) procedures on a set of real and simulated data, and we find the following. First, stellar kinematics cannot be a dominant source of error or bias since we find that a systematic change of 10\% of measured velocity dispersion leads to only a 0.7\% shift on \Hc from the seven lenses analyzed by TDCOSMO. Second, we find no bias to arise from incorrect estimation of the line-of-sight effects. Third, we show that elliptical composite (stars + dark matter halo), power-law, and cored power-law mass profiles have the flexibility to yield a broad range in \Hc values. However, the TDCOSMO procedures that model the data with both composite and power-law mass profiles are informative. If the models agree, as we observe in real systems owing to the "bulge-halo" conspiracy, \Hc is recovered precisely and accurately by both models. If the two models disagree, as in the case of some pathological models illustrated here, the TDCOSMO procedure either discriminates between them through the goodness of fit, or it accounts for the discrepancy in the final error bars provided by the analysis. This conclusion is consistent with a reanalysis of six of the TDCOSMO (real) lenses: the composite model yields \Hc=$74.0^{+1.7}_{-1.8}$ \ksmpc, while the power-law model yields $74.2^{+1.6}_{-1.6}$ \ksmpc. In conclusion, we find no evidence of bias or errors larger than the current statistical uncertainties reported by TDCOSMO. 

}

\titlerunning{Uncertainties in time-delay cosmography}
\maketitle

%
%-------------------------------------------------------------------
\section{Introduction}
\label{sec:introduction}
The time-delay method applied to gravitationally lensed quasars \citep{Refsdal1964} provides a perhaps unrivalled combination of high sensitivity to the Hubble constant \Hc, and minimal dependence on the other cosmological parameters, while relying only on well known physics (i.e., gravity). These qualities make this method particularly important in the present context, where there is growing evidence for tension in \Hc measurements using cosmological probes based on the early Universe and the late Universe \citep{Verde2019}. The power of the method in providing reliable \Hc measurements depends on three main factors: 1- precise time-delay measurements between multiple images of the background source, 2- well constrained models of the dominant primary and nearby lens galaxies, and 3- an estimate of the combined lensing effect of all the mass along the line of sight up to the redshift of the lensed quasar.  

Precise and accurate time-delay measurements are available, for example, from the COSMOGRAIL collaboration, using long-term photometric monitoring of selected lensed quasars \citep[e.g.,][]{Courbin2018, Bonvin2018, Bonvin2019}. The precision and accuracy of the COSMOGRAIL technique have been verified via a blind time-delay challenge \citep{Dobler2015,Liao2015,Bonvin2016}. The time-delays were then used to constrain cosmological parameters with detailed modeling of the potential well of the lens using the constraining power of sharp Hubble Space Telescope (HST) images \cite[e.g.,][]{Suyu2010, Suyu2014, Wong2017, Birrer2019, Rusu2019} or Keck AO imaging \cite[e.g.,][]{Chen2019}. The measured stellar kinematics of the lensing galaxy were used to mitigate the impact of well-known lensing degeneracies on the cosmological inference \citep[e.g.,][]{Treu2002}. Finally, multi-band wide-field imaging and/or spectroscopy \cite[e.g.,][]{Rusu2017, Sluse2019} was used to constrain the combined lensing effect of the line-of-sight objects and large-scale structures in a statistical way \citep{Greene2013,Rusu2017}. \cite{Tihhonova2018} also show that these estimates of the line-of-sight effects are compatible with the ones obtained with weak gravitational lensing.

Adopting these data and methodology, the H0LiCOW collaboration \citep{Suyu2017} is analyzing a sample of lenses suitable for high-precision \Hc measurements. The latest results based on six systems are summarized by \citet[][]{Wong2019}. We stress that the H0LiCOW results are obtained through blind analyses, in the sense that the mean value of all the observed cosmological parameters is hidden to the investigators until the analysis is complete and the papers have been written\footnote{The first lens system analyzed using the then newly developed lens modeling methods was not blinded (\Bsixteen), but the subsequent analyses of the other five lenses using similar methods were blinded.}. The goal of this procedure is to avoid conscious or unconscious bias from the experimenters.
We note that the six measurements that have been published thus far are statistically consistent with each other, in the sense that the scatter between the measurements is as expected from the estimated uncertainties. This means that if there are any unknown uncorrelated sources of error, those are subdominant with respect to the ones currently considered.

The resulting value of the Hubble constant in a flat $\Lambda$CDM universe, $\Hc = 73.3^{+1.7}_{-1.8}$ \ksmpc (2.4\% precision), is 3$\sigma$ higher than the early-Universe results \citep{Planck2018}, adopting the same \lcdm cosmological model, and is in very good agreement with other independent local measurements \citep[e.g.,][]{Riess2019}.  When combined with completely independent results from other local measurements of \Hc, the tension with the early-Universe probes range between 4 and 6$\sigma$ \citep{Verde2019}, depending on the combination of probes. Very recently, \cite{Pandey2019} also carried out statistical tests independent of any underlying cosmology, showing that the distances measured with strong lensing time delays and with supernovae, which are both local but independent measurements, are fully compatible \citep[see also][]{WojtakAgnello2019}. Although they cannot exclude that supernovae and lenses share exactly the same systematics, these systematic biases would also have to be preserved across redshift, which seems unlikely.

The blind analysis of a seventh lens system using very similar methods for the lens modeling, time-delay measurement, external convergence estimation and kinematics modeling to those adopted by H0LiCOW has recently been published by the STRIDES collaboration \citep{Shajib2019}. This work finds 74.2$^{+2.7}_{-3.0}$ \ksmpc, in agreement with the H0LiCOW result (an independent analysis adopting a different modeling software is currently under way). This most recent system is particularly interesting since it has two sets of multiple images at different redshifts, which help break some of the degeneracies, and results in the most precise individual measurement so far. In order to make further progress in this important area, members of the COSMOGRAIL, H0LiCOW, SHARP and STRIDES collaborations interested in time-delay cosmography of lensed quasars have decided to join forces with other scientists and form a new "umbrella" collaboration named TDCOSMO\footnote{\url{www.tdcosmo.org}} (Time-Delay COSMOgraphy).

The high statistical significance of the tension between early and late Universe probes has prompted two lines of investigation. On the one hand, theorists have been trying to find ways to reconcile the measurements by considering models beyond the standard $\Lambda$CDM one \citep[e.g.,][]{Knox2019}. On the other hand, observing teams have been focusing on increasing the precision of each method while carrying out tests of potential systematic uncertainties to ensure that the tension is real. After all, \textit{"extraordinary claims require extraordinary evidence".}

In this work, the first by the TDCOSMO collaboration, we explore a number of potential systematic uncertainties that may affect the time-delay cosmography method, after reviewing its methodology and implementation by TDCOSMO in Section~\ref{sec:msd} and the inference procedure in Section~\ref{sec:toysmodels}.
First, in Section~\ref{sec:kinem_sensitivity} we explore potential biases introduced by systematic uncertainties in the modeling and measurement of the deflector stellar velocity dispersion. Second, in Section~\ref{sec:corr}, we study uncertainties in the modeling of the line-of-sight contribution. Third, in Section~\ref{sec:6}, we address the long standing issue of the mass-sheet degeneracy and the flexibility of lensing models. It is very well known that assumptions must be made on the form of the main deflector mass distribution to break the mass-sheet degeneracy. As many authors have pointed out \citep{Falco1985,Read2007,Schneider2013,Xu2016,Sonnenfeld2018,CSK19}, if the models adopted are insufficiently flexible, the resulting uncertainties are underestimated and potentially biased. Section~\ref{conclusion} offers a summary and conclusions.

We address these three sources of potential systematic uncertainties using a combination of observational tests and simulations. We stress that a full simulation of the observational setup and lens modeling procedure is needed if one wants to obtain quantitative estimates of the uncertainties. Previous works \citep{Schneider2013,Sonnenfeld2018,CSK19} were based on idealized, often spherical models. Those are useful to gain intuition of the problem, but by their very nature cannot provide quantitative estimates due to the extreme approximation and the limited information utilized to constrain them, often just the Einstein Radius and an integrated velocity dispersion. The only way to obtain a faithful estimate of the uncertainties is to reproduce the measurement using the same amount of information (thousands of pixels from imaging, multiple time-delays, stellar kinematics) and modeling techniques. 
%The simulated dataset shown in this paper are carried out using the pipeline developed by \citet{Ding2017a,Ding2017b} and \citet{Ding2018}, and the fitting procedure mimics as closely as possible that of the H0LiCOW/SHARP/STRIDES (hereafter TDCOSMO) collaboration.
The simulated dataset shown in this paper are produced using the pipeline developed by \citet{Ding2017a,Ding2017b} and \citet{Ding2018}. In order to isolate and quantify the uncertainties associated with the lens mass modeling procedure, the simulated data consist of high-resolution images of lens systems comparable to the real observations, high-precision time delays (higher precision than those of real lenses so that the time-delay uncertainties are subdominant compared to the modeling uncertainties that we aim to quantify), and do not include the line-of-sight structures.  The lens fitting procedure that we use to analyze these simulated data resembles as closely as possible that of the TDCOSMO collaboration.

\section{Background \label{sec:msd}}
\subsection{Time-delay cosmography and the mass-sheet degeneracy}

Time delays in gravitationally lensed quasars provide a direct measurement of the so-called ``time-delay distance'', which is a combination of angular diameter distances to the source, \Ds, to the deflector, \Dd, from the deflector to the source, \Dds, and the redshift of the deflector \zd:
\begin{equation}
    \label{eq:ddt}
    \Ddt \equiv (1+\zd)\frac{\Dd \Ds}{\Dds}
\end{equation}
\citep{Refsdal1964, Schneider1992, Suyu2010}.

This quantity is related to the relative time delay between two multiple images A and B, $\Delta t_{AB}$, by: 
\begin{equation}
    \label{eq:td}
    \Delta t_{AB} = \frac{\Ddt}{c}  \left[ \frac{(\vec{\theta_A} - \vec{\beta})^2}{2} - \frac{(\vec{\theta_B} - \vec{\beta})^2}{2} - \psi(\vec{\theta_A)} + \psi(\vec{\theta_B}) \right], 
\end{equation}
where $\vec{\theta}$ is the image position on the plane of the sky, $\vec{\beta}$ is the (unobservable) source position, $c$ is the speed of light and $\psi$ is the lensing potential which is defined such that the deflection angle $\vec{\alpha}(\vec{\theta})$ is given by $\vec{\alpha}(\vec{\theta}) \equiv \nabla \psi(\vec{\theta})$. From Equation (\ref{eq:td}), we see that \Ddt depends on the geometry of the lensed system and on the potential well of the lensing galaxy. The mass profile is expressed as a dimensionless surface mass density, $\kappa(\vec{\theta})$, called the convergence. It is related to how the light beams from the source are stretched or squeezed, leading to an apparent (de)magnification and can be expressed as half of the Laplacian of the lensing potential: 
\begin{equation}
    \label{eq:kappadef}
    \kappa(\vec{\theta}) = \frac{1}{2}\nabla^2\psi(\vec{\theta}).
\end{equation}

We can also define the Fermat potential $\phi$ \citep{Schneider1985, Blandford1986} as 
\begin{equation}
    \label{eq:fermat}
    \phi(\vec{\theta}) \equiv \frac{(\vec{\theta} - \vec{\beta})^2}{2} - \psi(\vec{\theta}).
\end{equation}
Using this definition, Equation (\ref{eq:td}) reduces to
\begin{equation}
    \label{eq:fermat-td}
     \Delta t_{AB} = \frac{\Ddt}{c}\left[\phi(\vec{\theta_A}) - \phi(\vec{\theta_B})\right] \equiv  \frac{\Ddt}{c} \Delta \phi_{AB},
\end{equation}
where $\Delta \phi_{AB}$ is the difference of Fermat potentials at the positions of the multiple images.  
Based on the multipole decomposition of the gravitational potential, \cite{Kochanek2002} shows that the time-delay distance, \Ddt, depends on the mean surface density $\langle\kappa\rangle$ at the Einstein radius $\theta_{\mathrm{E}}$, specifically over the annulus defined by image positions.  

An inherent limitation of the lensing models to infer \Ddt is the so called Mass-Sheet Transformation \citep[MST, e.g.,][]{Falco1985} and its generalization \citep{Saha2000, Saha2006, Liesenborgs2012, Schneider2014, Wagner2018, Wertz2018}. The MST transforms the projected mass distribution and the source plane position according to:
\begin{equation}
\begin{aligned}
&\kappa(\theta) \rightarrow \kappa_{\lambda}(\theta) = (1 - \lambda) \times \kappa(\theta) + \lambda, \\ 
&\beta \rightarrow \beta' = \lambda\beta,
\end{aligned}{}
\end{equation}
where $\beta$ is the (unknown) source position on the sky prior to lensing. In other words, one can add a mass sheet to any model and apply a scaling factor, $\lambda$, without changing the lensing observables except the time delays and therefore the inferred cosmology. 

The time-delay distance given by any model is affected by MST as follows : 
\begin{equation}
    \label{eq:MSD}
    \Ddt^\mathrm{MST} = \Ddt^\mathrm{true}\times (1 - \lambda). 
\end{equation}
In the TDCOSMO analyses, this scaling factor $\lambda$ is identified with the external convergence factor \kext which accounts for the contribution of all the mass along the line of sight (LOS). It is estimated independently from the lens modeling by comparing the relative number of galaxies weighted by physically relevant priors such as the distance to the lens, the stellar mass and the redshift in a large aperture around the strong lens system with simulated LOS extracted from numerical simulations with similar statistical properties \citep{Rusu2017}. Alternatively, the external convergence can be estimated from a weak lensing analysis \citep{Tihhonova2018}.

%\shscom{Added paragraph below to address Dominique's comment}
In addition to the MST above due to external mass sheets (i.e., external mass structures that do not affect the stellar dynamics of the foreground lens galaxy), MST can also manifest itself approximately as a change in the radial mass profile of the foreground lens galaxy.  We describe this as an "internal" mass sheet.  To mitigate the effects of the internal mass sheet, we consider different families of models and further use kinematic measurements of the foreground lens that provide additional constraints on the lens mass models. In particular, the goodness of fit to the kinematic data, especially spatially-resolved lens stellar velocity dispersion, allows us to distinguish between otherwise degenerate lensing mass models \citep[e.g.,][]{Yildirim2019}.

The lens stellar velocity dispersion of the foreground lens galaxy allow the inference of the angular diameter distance, \Dd, to the lens, in addition to the time-delay distance \citep{ParaficzHjorth2009, Jee2015, Jee19}.  The inference of \Dd depends on the anisotropy of stellar orbits \citep{Jee2015}, but this additional distance measurement provides more leverage on constraining cosmological models \citep{Jee2016, Shajib2018}.  

\subsection{Two-distance inference}

In the most recent analysis of \Jtwelve, \PGeleven, \RXJ, \Bsixteen and \DESzerofour \citep{Birrer2016, Birrer2019, Chen2019, Shajib2019, Wong2019}, the time-delay distance \Ddt and the angular diameter distance to the lens \Dd are jointly inferred. Following the method developed in \cite{Birrer2016}, the luminosity weighted LOS velocity dispersion within an aperture $\mathcal{A}$ of the main deflector $\sigma_v$ can be expressed as: 
\begin{equation}
    \label{eq:sigma-Dd}
    \sigma_v ^2 = (1 - \kext) \frac{\Ds}{\Dds}c^2J(\xi_\mathrm{lens}, \xi_\mathrm{light}, \beta_\mathrm{ani}), 
\end{equation}
where $\xi_\mathrm{lens}$ is the set of all parameters contained in the lens mass model, $\xi_\mathrm{light}$ is the parameter of the light models and $J$ is a function that captures all dependencies on the modeling parameters and the anisotropy profile $\beta_\mathrm{ani}$. Using Equation (\ref{eq:ddt}), (\ref{eq:fermat-td}) and (\ref{eq:MSD}), we have : 
\begin{equation}
    \label{eq:a}
    \frac{\Dd \Ds}{\Dds} = \frac{c \Delta t_{AB}}{(1+z_d)(1 - \kext)\Delta \phi_{AB}(\xi_\mathrm{lens})}.
\end{equation}

Combining Equations (\ref{eq:sigma-Dd}) and (\ref{eq:a}), we obtain an expression for the angular diameter distance to the lens which is independent of the external convergence: 
\begin{equation}
    \Dd = \frac{c^3 \ \Delta t_{AB} \ J(\xi_\mathrm{lens},\, \xi_\mathrm{light},\, \beta_\mathrm{ani})}{(1+\zd) \ \sigma_v^2 \ \Delta \phi_{AB}(\xi_\mathrm{lens})}.
\end{equation}
We immediately see that the angular diameter distance \Dd varies as $\frac{1}{\sigma_v^{2}}$. The dependence of \Dd to a change in the measurement of $\sigma_v$ can therefore be computed analytically : 
\begin{equation}
\label{eq:Ddvss}
    \frac{\mathrm{d}\,\Dd}{\Dd} = -2 \frac{\mathrm{d}\,\sigma_v}{\sigma_v}, 
\end{equation}
whereas \Ddt is left unchanged when varying the velocity dispersion. 
The final \Hc measurement is obtained by combining these two distance measurements. As a consequence, the importance of the velocity dispersion in the final \Hc value depends on the relative precision between the angular diameter distance and the time-delay distance, and on the mapping between the parameters. The \Ddt measurement is typically more constraining of \Hc than \Dd given the current observational data. Future observations with spatially resolved kinematics are expected to improve substantially the \Dd constraints \citep{Yildirim2019}.

Two of the lens systems in the TDCOSMO sample, \HEzero\ and \WFItwenty, have nearby massive perturbing galaxies at a different redshift from the strong lensing galaxy, and thus required multi-lens-plane mass modeling.  The single-lens-plane equations (\ref{eq:sigma-Dd})-(\ref{eq:a}) are thus not directly applicable, given the additional angular diameter distances involved in the multiple lens planes.  Nonetheless, the mass model of the lens galaxy can still be used to predict the velocity dispersion to compare to the measured value, so the kinematic measurement can be used to further constrain the mass model.  It turns out that an effective time-delay distance could be derived for these two lens systems, but the inference of \Dd accounting for the multi-lens planes is deferred to future work.

%%%%%%%%%%%%%%
\subsection{The current TDCOSMO model families}
\label{sec:current_model_familly}
%%%%%%%%%%%%%%

The collaborations within TDCOSMO currently consider two classes of models (composite and power-law), to reconstruct the mass distribution of the main lens, with the exception of the first system analyzed \Bsixteen \citep{Koopmans2003, Suyu2010}. \Bsixteen was modeled only using a power-law, as \cite{Suyu2009} showed that deviations to a smooth potential using pixellated corrections were negligible. The fact that the corrections are so small, even though the deflector in this complex lens is an obvious merger between two galaxies, is a remarkable indication of the degree of smoothness of the overall gravitational potential. This is also supported by the analysis of extended rings used to detect substructures in lenses through their impact on the smoothness of Einstein rings. Aside from specific features arising from well-identified substructures in any given lens, no statistically significant correction to simple parametric lens models is found by \citet{Vegetti14}.

For the above reasons, the TDCOSMO analyses consider purely analytical lens models with sufficient degrees of freedom to catch a broad range of observables given current imaging capabilities with HST or adaptive optics. More specifically, the TDCOSMO analyses considers elliptical power-law and composite models, with the addition of external shear.

\subsubsection{Power-law model  \label{ssec:powerlaw}}
Power-law models have a constant projected mass slope over the entire profile. The convergence of the power-law elliptical mass distribution \citep{Barkana1998} is described by : 
\begin{equation}
\label{eq:powerlaw}
    \kappa_{\mathrm{PL}}(\theta_1, \theta_2) = \frac{3 - \gamma}{2} \left[ \frac{\theta_{\mathrm{E}}}{\sqrt{q_\mathrm{m}\theta_1^2 + \theta_2^2/q_\mathrm{m}}}\right] ^{\gamma - 1}, 
\end{equation}
where $\gamma$ is the slope of the profile, $q_m$ is the axis ratio of the elliptical profile and $\theta_{\mathrm{E}}$ is the Einstein radius. The coordinate system is defined such that the $\theta_1$ and $\theta_2$ coordinates are along the major and minor axis respectively. The cored power-law profile is a natural extension of this model which introduces an additional free parameter, namely the core radius in the center of the profile $\theta_{\rm c}$ and is defined as : 
\begin{equation}
    \kappa_{\mathrm{cPL}}(\theta_1, \theta_2) = \frac{3 - \gamma}{2} \left[ \frac{\theta_{\mathrm{E}}}{\sqrt{q_\mathrm{m}\theta_1^2 + \theta_2^2/q_\mathrm{m} + \theta_{\rm c}^2}}\right] ^{\gamma - 1}.
\end{equation}
This profile has therefore a shallower slope in the center to reproduce the core of galaxies. A complete description of this mass model can be found in \cite{Barkana1998}. Although not used by the TDCOSMO collaboration, except in the analysis of \RXJ by \citet{Suyu2014} who found negligible core size, we tested cored power-law profiles on simulated lenses in Section~\ref{sec:6}. 

\subsubsection{Composite model \label{ssec:composite}}
The second family of mass models used by the TDCOSMO collaboration are the so-called composite models, which consist of baryonic matter and dark matter components. For the dark matter, a Navarro-Frenk-White (NFW) profile is used. The spherical NFW density distribution is given by : 
\begin{equation}
\label{eq:nfw}
    \rho_{\mathrm{NFW}}(r) = \frac{\rho_s}{(r/r_s)(1+r/r_s)^2},
\end{equation}
where $r_s$ is the scale radius and $\rho_s$ is a normalization factor \citep{Navarro1997}. For the baryonic component, the TDCOSMO collaboration adopts the Chameleon profile, which is the difference between two singular isothermal ellipsoids and closely mimics a S\'ersic profile. A complete description of this model can be found in \cite{Dutton2011} and \cite{Suyu2014}. This family of mass model allows more flexible mass distribution than power-law models since the slope of the projected mass profile is not constant over the whole lens galaxy. 

%%%%%%%%%%%%%%
\section{Inference procedure and limitations of toy models}
\label{sec:toysmodels}
%%%%%%%%%%%%%%

The next step required to derive a \Hc measurement from the data is a statistical inference. The collaborations contributing to TDCOSMO adopt a Bayesian framework and compute the posterior probability distribution function of all the cosmological and nuisance parameters given the data.

The imaging and spectroscopic data contain huge amounts of information, well beyond the position of the quasar images. Setting aside the line of sight, which is constrained independently, the main sources of constraints for the main deflector(s) mass models are: the pixels of the high resolution images (of order 10$^4$); independent time delays (up to three for a quad); stellar velocity dispersion of the main deflector and nearby perturbers, if present. The inference required to extract all the information from the data is computationally very intensive. Taking into account the need to explore multiple and flexible models to marginalize over modeling choices, the TDCOSMO analysis required up to a million CPU hours per lens.

In the recent past, simplified toy models, that is, models in which either i) the lens systems are not simulated with sufficient complexity, or ii) the inference procedure does not exploit the full information content, have been used to investigate systematic uncertainties in time-delay cosmography \citep{Schneider2013,Sonnenfeld2018,CSK19}. These models are certainly a useful illustration, and it is encouraging that they conclude that a precision within the range 3-10\% can be reached with their simplified approach and limited constraints. However, owing to their limitations, those models cannot provide the quantitative answers that are needed to understand whether there are biases at the 2\% level, which is the current achievement of time-delay cosmography. Chief among the limitations of previous works is the use of spherical models. Spherical models are inherently inappropriate to model quads \citep[e.g.,][]{Kochanek2006}, because they cannot even produce four images and thus are intrinsically less constrained by the data than observed quads. 

The bulk of the lensing information comes from the radial extent and surface brightness distribution of the lensed images, which constrains directly the radial dependency of the mass distribution, the key parameter driving the inference of \Hc. Toy models neglect this information \citep[e.g.,][]{CSK19}, and are mostly spherical and constrained solely by the position of the quasar images spanning just 10\% on either side of the Einstein radius. Furthermore, they are constrained only by the positions of the multiple images of the quasars and not using the full information content of the lensed host galaxy, often amounting to thousands of high signal-to-noise ratio pixels (see Section~\ref{sec:6} and Appendix~\ref{app:simple_model} for details). 
These constraints would have no way to detect significant departures from a power law for example, which could instead be detected in real-life cases as variations in the distortion of the images spanning a much larger significant radial range. Indeed, most of the HST data used in time-delay cosmography display prominent Einstein rings, spanning several tenths of arcseconds radially. In other words, the radial width of the ring is significant compared with the Einstein radius itself, hence constraining the potential well radially. This is clearly illustrated with the case of \RXJ\ in, for example, \cite{Suyu2014}. In addition, toy models typically condense the information in a few parameters and thus cannot realistically explore the degeneracies between true model parameters and how uncertainties in the actual data translate into inference.

\begin{figure*}[t!]
\centering
    \includegraphics[width=14cm]{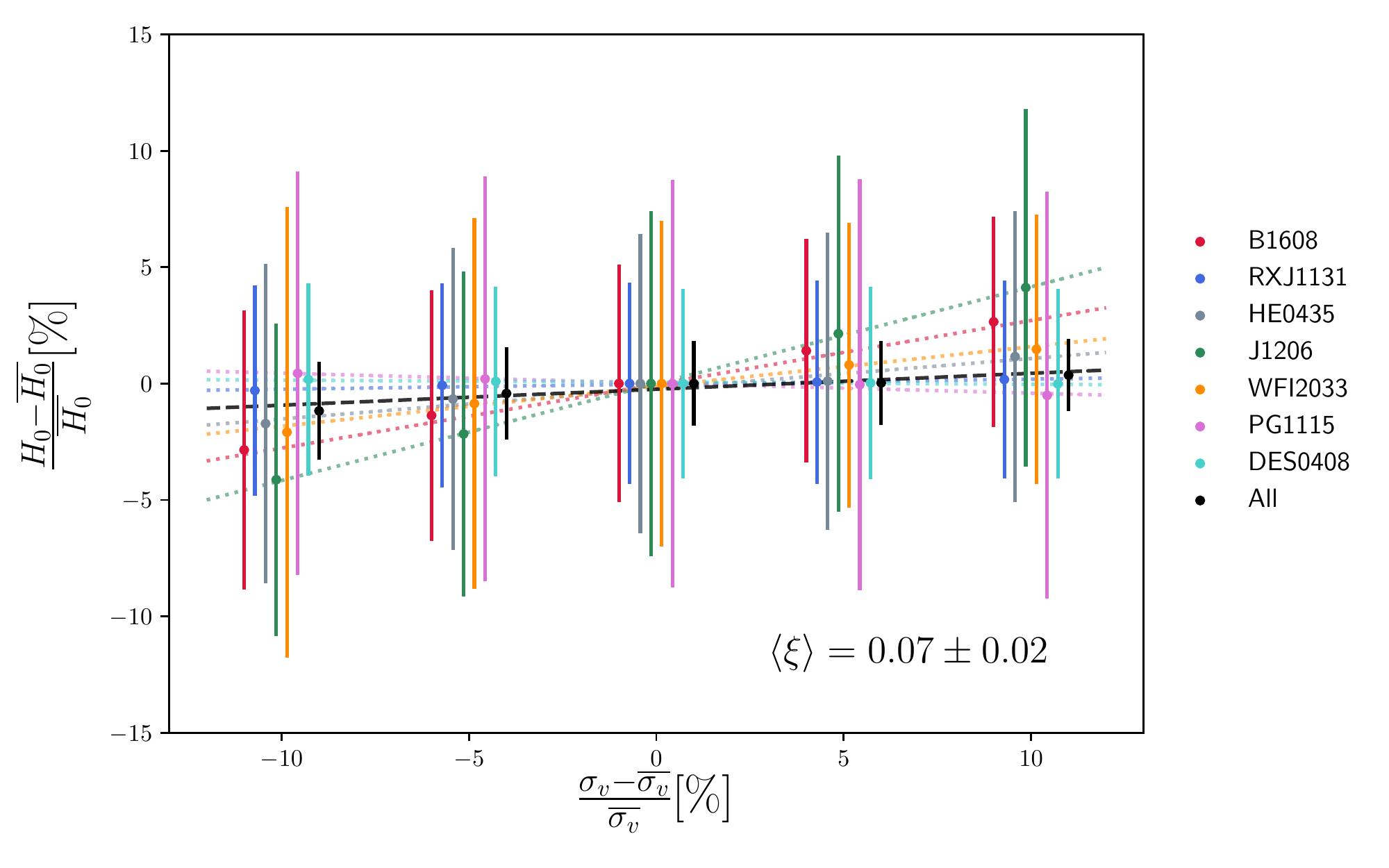}
    \includegraphics[width=0.49\textwidth]{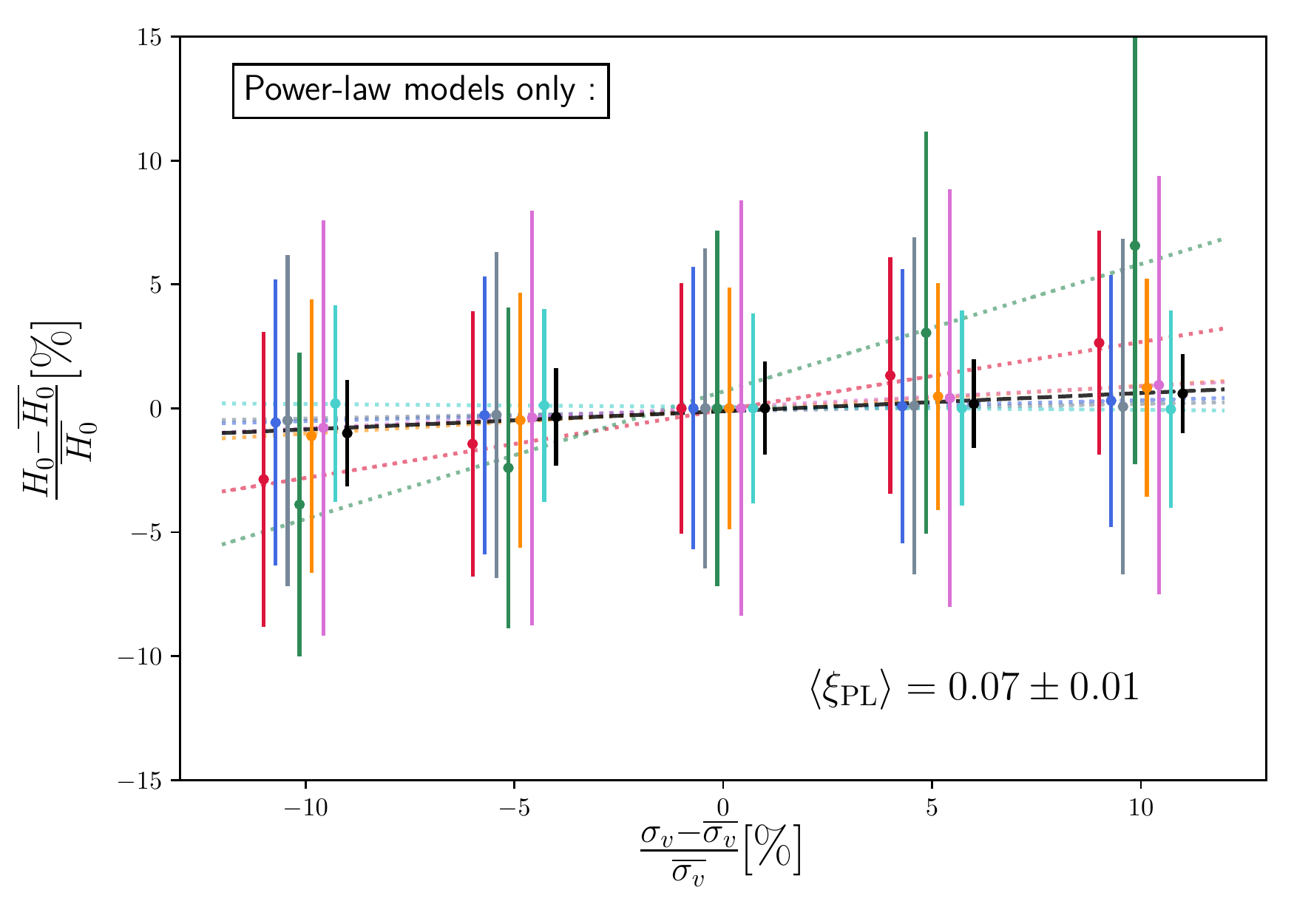}
    \includegraphics[width=0.49\textwidth]{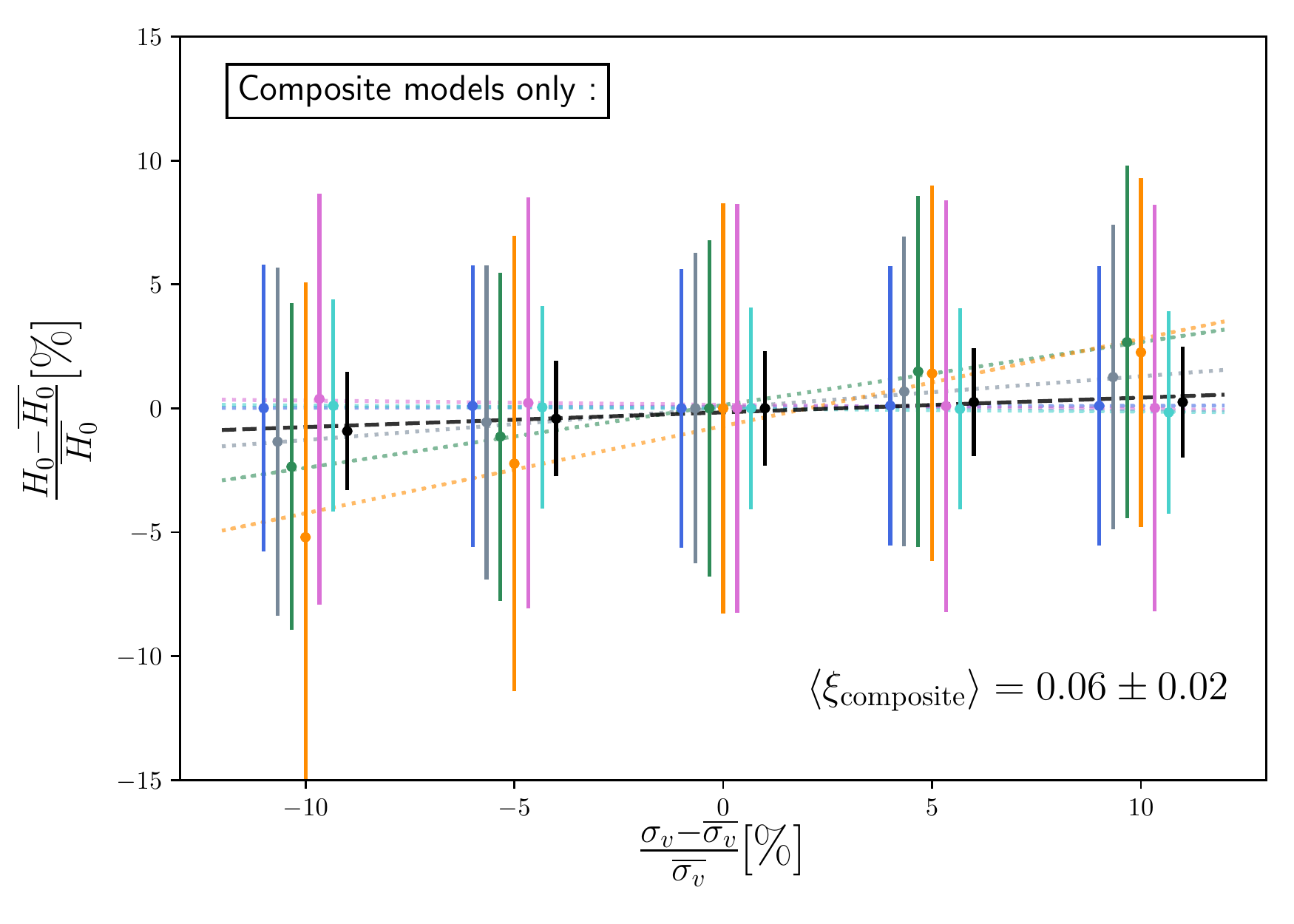}
    \caption{Sensitivity of the inferred Hubble constant as a function of fractional change in the measured lens velocity dispersion, $\sigma_v$ (see Eq.~\ref{eq:sensitivity}). Each color corresponds to one of the seven strong lens systems of the current TDCOSMO sample. The dotted lines display the best linear fit to the data. The joint inference performed on the seven lenses is shown in black. The error bars correspond to the $16^{th}$ and $84^{th}$ percentile of the posterior distributions. The two bottom panels show the sensitivity of \Hc to a change in the measured lens velocity dispersion for power-law (left) and composite (right) models independently. The sensitivity of the joint inference, $\langle \xi \rangle$ is indicated on each panel.} 
    \label{fig:h0vssigma}
\end{figure*}

Last but not least, it is crucial to have the ability to assess the goodness of the models, in both absolute and relative terms. This is to our knowledge the most powerful way to establish whether the chosen parametrization is an appropriate description of the data. The power of goodness of fit estimates depends on the realism and information content of the models. Models that are based on key summary statistics are able to use goodness of fit only on those statistics. More realistic models that produce and fit, for example, image positions will have a few more observables to assess goodness of it. Models that produce the full surface brightness distribution of the lenses source(s) and other observables will have access to many more observables, and thus have significantly larger power for model exploration and selection.

\section{Influence of kinematics data on \Hc measurement \label{sec:kinematics_impact}}
\label{sec:kinem_sensitivity}

One important ingredient to mitigate the impact of the MST is to use the kinematics of the deflector as an independent mass estimator \citep{Treu2002, Koopmans2004}, since within a cosmological model \Dd and \Ddt are related to each other. So far, the central stellar velocity dispersion integrated within an aperture, $\sigma_v$, has been used even though additional and substantial gains can be obtained by including spatially resolved information that helps break the mass-anisotropy degeneracy \citep{Barnabe2011,Czoske2012,Shajib2018,Yildirim2019}. 

The inference of the Hubble constant is driven by a combination of observables, including the extended images used in the lens model, multiple time delays if available, and kinematic information. Thus, the dependency of \Hc on kinematics data defined by
\begin{equation}
\label{eq:sensitivity}
    \xi \equiv \frac{\delta H_0 / \overline{H_0}}{\delta \sigma_v / \overline{\sigma_v}}
\end{equation}
cannot be estimated with simple dimensional arguments or toy models, but needs to be computed by repeating the inference while varying the input kinematics data. The result will depend on the details of the analysis as well as on the relative quality and constraining power of the kinematic and nonkinematic data, and on how the \Ddt $-$\,\Dd plane maps into \Hc as a result of the deflector and source redshifts. Each of these factors varies from lens to lens as we show below and thus cannot be simply derived from a toy model and generalized to every lens.

\subsection{The TDCOSMO analysis and its sensitivity to the measured velocity dispersion }

Simple models such as the Singular Isothermal Sphere (SIS) models, can have a very strong dependency on the velocity dispersion. This dependency could be on the order of $\xi \sim 1$, which means that a 1\% change in the velocity dispersion $\sigma_v$ leads roughly to a 1\% change in \Hc. The high sensitivity of SIS mass models to a change in the velocity dispersion arises from the fact that they have only one free parameter (the normalization). If galaxies were all SIS, then such a high sensitivity would allow us to better constrain the mass model through more precise and accurate kinematic measurements.

In this section we show that the TDCOSMO measurements, which use models more flexible than SIS and constrain them with a wealth of data, are less sensitive to the kinematics information than SIS. In order to quantify how the error on $\sigma_v$ propagates into \Hc, we recomputed the posterior distributions for \Ddt and \Dd after changing arbitrarily the median value for our $\sigma_v$ distribution. We perform the test for four values of the shift, that is, $\delta \sigma_v/\sigma_v = \pm$ 5\% and $\delta \sigma_v/\sigma_v = \pm$ 10\%, for each individual lens in the TDCOSMO sample as well as for the joint \Hc inference. Throughout this section, the \Hc inference was performed in flat \lcdm cosmology with a uniform prior on $\Omega_{\mathrm{m}} \in [0.05, 0.5]$.

Figure \ref{fig:h0vssigma} summarizes the results, where we define $\overline{H_0}$ and $\overline{\sigma_v}$ as the inferred \Hc value of the system and its measured aperture velocity dispersion. The models used in Fig.~\ref{fig:h0vssigma} include both composite and power-law mass models\footnote{The first H0LiCOW lens, namely \Bsixteen, was modeled with a power-law model and pixellated potential corrections, which were found to be small. A composite model was not applied, so we use only the power-law model in our analysis \citep[see][for details]{Suyu2010}. } combined according to the standard procedure described in previous papers \citep[e.g.,][]{Suyu2014, Chen2019, Birrer2019, Rusu2019}.  We first discuss in this section the general trend between $\sigma_v$ and \Hc for the combination of the two model families. Then, we discuss the specifics of each model family separately.

\begin{table*}[htbp!]
\centering
\renewcommand{\arraystretch}{1.3}
\fontsize{8.5}{12}\selectfont
\begin{tabular}{l|ccccccccc}
& \multicolumn{1}{c}{\begin{tabular}[c]{@{}c@{}}$H_0$\\ $[$\ksmpc$]$ \end{tabular}} & \multicolumn{1}{c}{\begin{tabular}[c]{@{}c@{}}$\sigma_v $\\ $[$ km.s$^{-1}]$\end{tabular}} & \multicolumn{1}{c}{\begin{tabular}[c]{@{}c@{}}\slope \\ (All models)\end{tabular}} & \multicolumn{1}{c}{\begin{tabular}[c]{@{}c@{}}$\xi_{\mathrm{PL}}$ \\ (power-law)\end{tabular}} & \multicolumn{1}{c}{\begin{tabular}[c]{@{}c@{}}$\xi_{\mathrm{composite}}$ \\ (composite)\end{tabular}} & Aperture  & \reff & \thetaE & \raper \\ \hline
\Bsixteen  & $71.0 ^{+2.9}_{-3.3}$  &   $260 \pm 15$      & $0.27 \pm0.01$      &   $0.27 \pm0.01$                 &  -    &$1\farcs00 \times 0\farcs84$  & $0\farcs59$ & $0\farcs81$ & $0\farcs46$  \\
\RXJ       & $78.2 ^{+3.4}_{-3.4}$  &   $323 \pm 20$        & $0.02\pm0.01$      & $0.04\pm0.01$            &   $0.01\pm0.01$     &$0\farcs81 \times 0\farcs70$  & $1\farcs85$ & $1\farcs63$ & $0\farcs38$ \\
\HEzero    & $71.7 ^{+4.8}_{-4.5}$  &   $222 \pm 15$      & $0.08 \pm0.01$      & $0.03\pm0.01$       &      $0.13\pm0.01$       &$0\farcs74 \times 0\farcs54$  & $1\farcs33$ & $1\farcs22$ & $0\farcs32$ \\
\Jtwelve   & $68.9 ^{+5.4}_{-5.1}$  &   $290 \pm 30$      & $0.42 \pm0.01$    & $0.51\pm0.06$   &       $0.25\pm0.01$         &$1\farcs90 \times 1\farcs00$     & $0\farcs34$ & $1\farcs25$ & $0\farcs73$\\
\WFItwenty & $71.6 ^{+3.8}_{-4.9}$  &   $250 \pm 19$        & $0.17 \pm0.02$      & $0.09\pm0.01$     &    $0.35\pm0.06$     & $1\farcs80 \times 1\farcs80$   & $1\farcs41$ & $0\farcs94$ & $0\farcs90$\\
\PGeleven  & $81.1 ^{+8.0}_{-7.1}$  &    $281 \pm 25$       & $-0.04 \pm0.01$     & $0.08\pm0.01$     &      $-0.02\pm0.01$      & $1\farcs06 \times 1\farcs00$ & $0\farcs53$ & $1\farcs08$ &  $0\farcs52$ \\
\DESzerofour & $74.2 ^{+2.7}_{-3.0}$ &  $227 \pm 9 $        & $-0.01 \pm0.01$            &    $-0.01\pm0.01$       &       $-0.01\pm0.01$        & $1\farcs00 \times 1\farcs00$ & $1\farcs20$ & $1\farcs92$ &  $0\farcs50$ \\
All        & -   & -                   & $0.07\pm0.02$               &   $0.07\pm0.01$      &     $0.06\pm0.02$         & -& -& -
\end{tabular}
\caption{\label{tab:slope} Summary of the $H_0$ values (Col.~2) reported in \cite{Wong2019} and \cite{Shajib2019}. Col.~3 gives the aperture velocity dispersion used for their analysis along with 1$\sigma$ error bars. Cols. 4-6 give the sensitivity, $\xi$, of the inferred $H_0$ value to the lens galaxy velocity dispersion. When the information is available, we make a distinction between composite and power-law model and the combination of these. Cols.~7-9 list the size of the aperture used for the velocity dispersion measurement, the effective radius \reff of the lens and the Einstein radius of each lens. Col.~10 give the aperture radius $\theta_{\mathrm{aperture}}$, computed by taking half of the average length of the slit side.}

\end{table*}

The slope \slope\ quantifies the sensitivity of the inferred \Hc value to a change in velocity dispersion. It is computed by performing a linear regression to the points (Table~\ref{tab:slope}). We observe large variation of measured slopes from object to object. However, for the full sample, the joint \Hc inference leads to a mean sensitivity of $\langle \xi \rangle = 0.07\pm0.02$. In other words, a systematic increase (decrease) of 10\% on the velocity dispersion increases (decreases) \Hc by approximately 0.7\%. 

\PGeleven and \DESzerofour differ from the other lenses with a slightly negative slope of \slope=$-0.04 \pm 0.01$ and \slope=$-0.01 \pm 0.01$ respectively. For the other lenses, increasing the velocity dispersion leads to a smaller angular diameter distance \Dd and therefore to a higher \Hc (Eq.~\ref{eq:Ddvss}). This behavior could be explained for \DESzerofour as this lens is a complex system with several sources located at two different redshifts. Thus, the reduced dependency on velocity dispersion could be due to the extraordinary azimuthal and radial extent of the lensing information, and the fact that multiple redshift sources might help limit the effects of MST. In this regime, the kinematics information only brings very limited constraints on the mass model. The measurement of \Hc is therefore almost insensitive to the kinematics.

In the case of \PGeleven, the time-delay distance \Ddt, which does not depend on the kinematics data, has a much larger constraining power on \Hc than the angular diameter distance \Dd. As a result, \PGeleven is also almost insensitive to the velocity dispersion. The same effect explains, to a lesser extend, the low sensitivity of \RXJ.
We note that \Jtwelve has the largest sensitivity to a change in $\sigma_v$, with an increase of 10\% in velocity dispersion leading to an increase of \Hc by 4.2\%. We interpret this as the effect of \Ddt being less well constrained by the lensing data on their own. The more limited lensing constraints with respect to other systems are probably because this is the only doubly imaged quasar in the sample - all the others are quadruply imaged. Last but not least, we note that \Jtwelve and \PGeleven have the largest relative uncertainty on $\sigma_v$ among the TDCOSMO sample. Therefore, the zero points on the x-axis of Fig.~\ref{fig:h0vssigma} for these two objects are the most uncertain.

\begin{figure*}[t!]
    \centering
    \begin{minipage}[c]{\textwidth}
    \includegraphics[width=\textwidth]{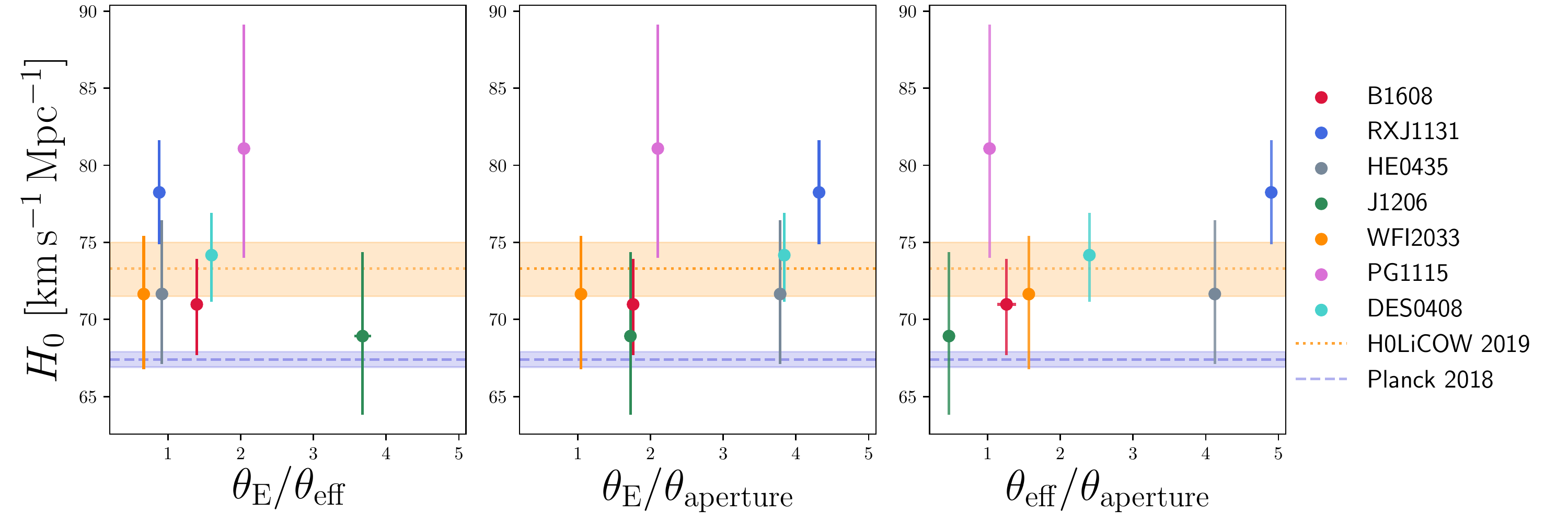}
    \end{minipage} 
    \caption{Effective radius \reff, Einstein radius \thetaE and radius of the spectroscopic aperture \raper of the TDCOSMO lenses. We show the ratios of these three quantities and the corresponding \Hc value inferred for each system. We do not observe significant correlations between the characteristic sizes of the lens, the spectroscopic aperture and \Hc. The horizontal lines indicate the latest H0LiCOW 2019 \citep[dotted orange, ][]{Wong2019} and Planck \citep[dashed blue, ][]{Planck2018} results along with the 1$\sigma$ uncertainties.}
    \label{fig:multiplot}
\end{figure*}

\subsection{Sensitivity to kinematics of the different mass model families}
We repeat the experiment for power-law models and composite model separately to check the sensitivity to kinematics data of each family of mass models. We do not use \Bsixteen when computing the sensitivity to kinematics of the composite models since this system has a pixelated potential correction performed on the power-law model, but no composite model. Bottom panels of Fig.~\ref{fig:h0vssigma} show the result of this test.

We obtain $\langle \xi_{\mathrm{composite}} \rangle=0.06\pm0.02$ and $\langle \xi_{\mathrm{PL}} \rangle=0.07\pm0.01$. The value of the joint inference is similar for both the composite and the power-law cases but each lens behaves differently. While \WFItwenty becomes more sensitive to the kinematics when modeled only with a composite model, \Jtwelve has its sensitivity almost halved. We can explain this behavior as due to the relative precision of the two families of models, which is different from one lens to the other. The time-delay distance of \Jtwelve is better constrained by composite models ($\Ddt = 5690^{+ 449}_{-356}$ Mpc at 7.1 \% precision) than with power-law models ($\Ddt = 5873^{+659}_{ -659}$ Mpc at 11.2 \% precision). The relative weight of the \Ddt compared to the \Dd in the final value of \Hc is therefore more important in the composite model case. 

\WFItwenty experiences the opposite behavior; it has tighter constrains with power-law models ($\Ddt = 4701^{+242}_{-204}$ Mpc at 4.74 \% precision) than with composite models ($\Ddt = 4909^{+485}_{-319}$ Mpc at 8.2\% precision). \WFItwenty is therefore more sensitive to the kinematics in the composite model case. 

In summary, there is no evidence that one family of mass models is significantly more sensitive to the kinematics than the other. For individual lenses, we observe differences but they can be explained by the relative precision that each of the models can achieve on the \Ddt measurement with respect to their \Dd measurement, based on the relative weight of the lensing and kinematic constraints and on the redshift of deflector and source that determine how the \Ddt\,-\,\Dd constraint maps into \Hc.

%%%%%%%%%%%%%%%%%%%%%%
\section{Search for correlations between \Hc and physically independent observables}
%%%%%%%%%%%%%%%%%%%%%%
\label{sec:corr}

The inference of \Hc relies on many independent ingredients and observables, such as the velocity dispersion of the deflector and the relative density of galaxies in the line of sight up to the background quasar. Those quantities do not have any physical reason to be correlated with \Hc. Thus, any evidence of a correlation between these observables and the inferred value of \Hc across the TDCOSMO sample, beyond the expected error covariance, would be an indication of underlying systematic errors. In this section, we carry out a number of empirical tests, correlating \Hc with observables and properties of the instrumental setup, and find no evidence for any statistically significant dependency.

\subsection{Dependency on the characteristic scale of the lens system and spectroscopic aperture. \label{ssec:radii_ratios}}

Figure \ref{fig:multiplot} shows the inferred Hubble constant for each of the seven TDCOSMO lenses for several combinations of characteristic scales of the lens systems and the aperture used for spectroscopic follow-up. In the left panel, we use the ratio between the Einstein and the effective radii to investigate any departure from the assumed description of the radial mass density profile. The ratio between the effective radius and the Einstein radius is used as a diagnostic of the relative spatial distribution of luminous and total matter. If the TDCOSMO models were insufficiently flexible, one may expect a trend in this ratio because the sum of the dark and luminous component would produce different shape of the total mass profile and a lack of flexibility in the mass model would not be able to reproduce the correct underlying distribution. In the middle panel are shown the ratios between Einstein radius and the spectroscopic aperture, which compare the spatial scales at which the lensing and kinematic information is obtained. Finally, the right panel of Fig.~\ref{fig:multiplot}, shows the ratio between the effective radius and the radius of the spectroscopic aperture, which could potentially be affected if the stellar kinematics were incorrectly modeled. One expects trends in all the above quantities if, for example, the assumptions about orbital anisotropy were systematically wrong. 

The Spearman's rank correlation coefficient between $H_0$ and $\theta_{\mathrm{E}}/\theta_{\mathrm{eff}}$, $\theta_{\mathrm{E}}/\theta_{\mathrm{aperture}}$ and $\theta_{\mathrm{eff}}/\theta_{\mathrm{aperture}}$ are respectively -0.11, 0.42 and 0.67. The probability that an uncorrelated data set produces such correlation coefficients (i.e., the p-value) is 0.82, 0.33 and 0.10. Therefore we conclude that in all three cases, there is no statistically significant correlation, even though the dynamical range on the $x$-axis is a factor of 3--6. While the absence of correlations does not prove that all systematic errors are below the statistical uncertainties, this is an important sanity check for our current models and for future work as the statistical precision improves with growing sample size.

In addition, observational and modeling effects such as the choice of stellar template, the choice of anisotropy model, or the PSF modeling could potentially bias the measured velocity dispersion of the main deflector and thus \Hc. The net effect of all these possible sources of systematic errors is difficult to quantify exactly but they typically scale with the effective radius of the lens \reff or the aperture radius of the spectroscopic observation \raper. The absence of any trend in Fig.~\ref{fig:multiplot} is reassuring in this regard. Moreover, as we showed in Section \ref{sec:kinem_sensitivity}, even $\sim 5\%$ systematic bias on the measured velocity dispersion, or equivalently on the modeled quantities due to incorrect anisotropy assumptions, will only produce an average 0.35\% bias on \Hc. Furthermore, as shown above, the direction and amplitude of the error would be different for each lens and therefore this systematic uncertainty would also show as a source of scatter or trend across the sample, which are not observed.

\begin{figure}[t!]
    \centering
    \begin{minipage}[c]{0.49\textwidth}
    \includegraphics[width=\textwidth]{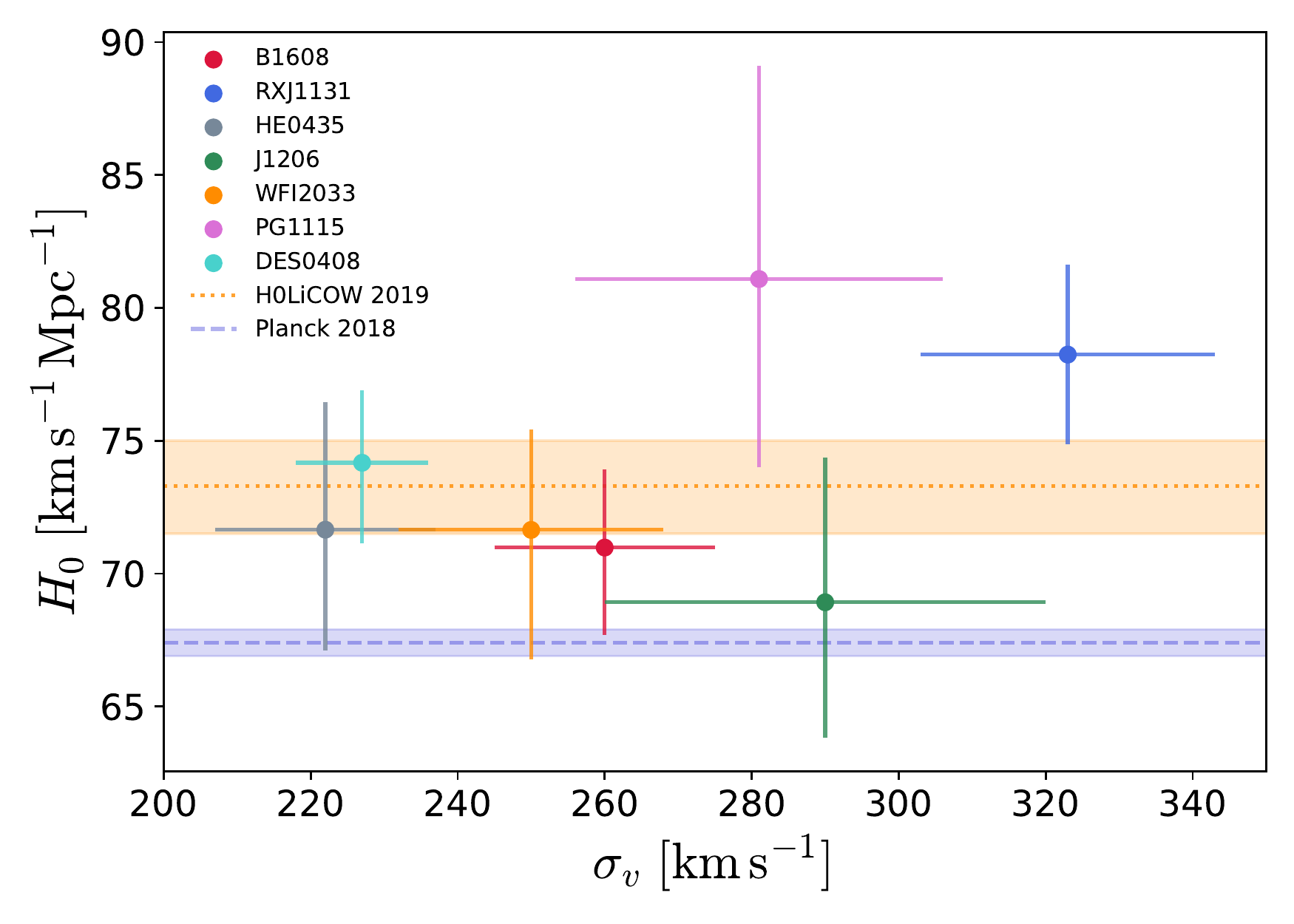}
    \end{minipage} 
    \caption{Hubble constant as a function of the measured velocity dispersion of the main lens. The horizontal lines indicate the latest H0LiCOW 2019 \citep[dotted orange, ][]{Wong2019} and Planck \citep[dashed blue, ][]{Planck2018} results along with the 1$\sigma$ uncertainties.}
    \label{fig:h0vsveldisp}
\end{figure}

\subsection{Dependency on intrinsic parameters of the deflector traced by the velocity dispersion}

An additional potential concern is whether systematic differences between our assumptions and the internal structure of early-type galaxies could give rise to measurable biases. For example, the so-called "tilt" of the fundamental mass plane is believed to arise primarily from the increase in dark-to-stellar matter ratio, a systematic change in stellar initial mass function with galaxy stellar mass, and possibly a small subdominant contribution from systematic variations in stellar orbits anisotropy \citep{Auger2010,Cappellari2016}. The stellar initial mass function is not a concern in the TDCOSMO analysis, since the stellar mass to light in the composite models is a free parameter. However, in principle the other two sources of "tilt" could introduce a potential systematic effect in TDCOSMO analysis, where each system is analyzed independently and with the same priors, rather than with priors that depend on the stellar mass.

In Fig.~\ref{fig:h0vsveldisp} we show the inferred \Hc as a function of stellar velocity dispersion, a redshift independent proxy of position along the fundamental plane. In this case, we found a Spearman's rank correlation coefficient of 0.07 with a p-value of 0.88. Hence, we conclude that there is no statistically significant trend in these data, indicating that any residual velocity dispersion dependent bias is smaller than the measurement uncertainties, and thus not significant at this stage.  As for the plots shown in the previous (and next) section, this sanity test should be repeated as the sample size and individual measurement precision increase.

\begin{figure}[t!]
    \centering
    \begin{minipage}[c]{0.49\textwidth}
    \includegraphics[width=\textwidth]{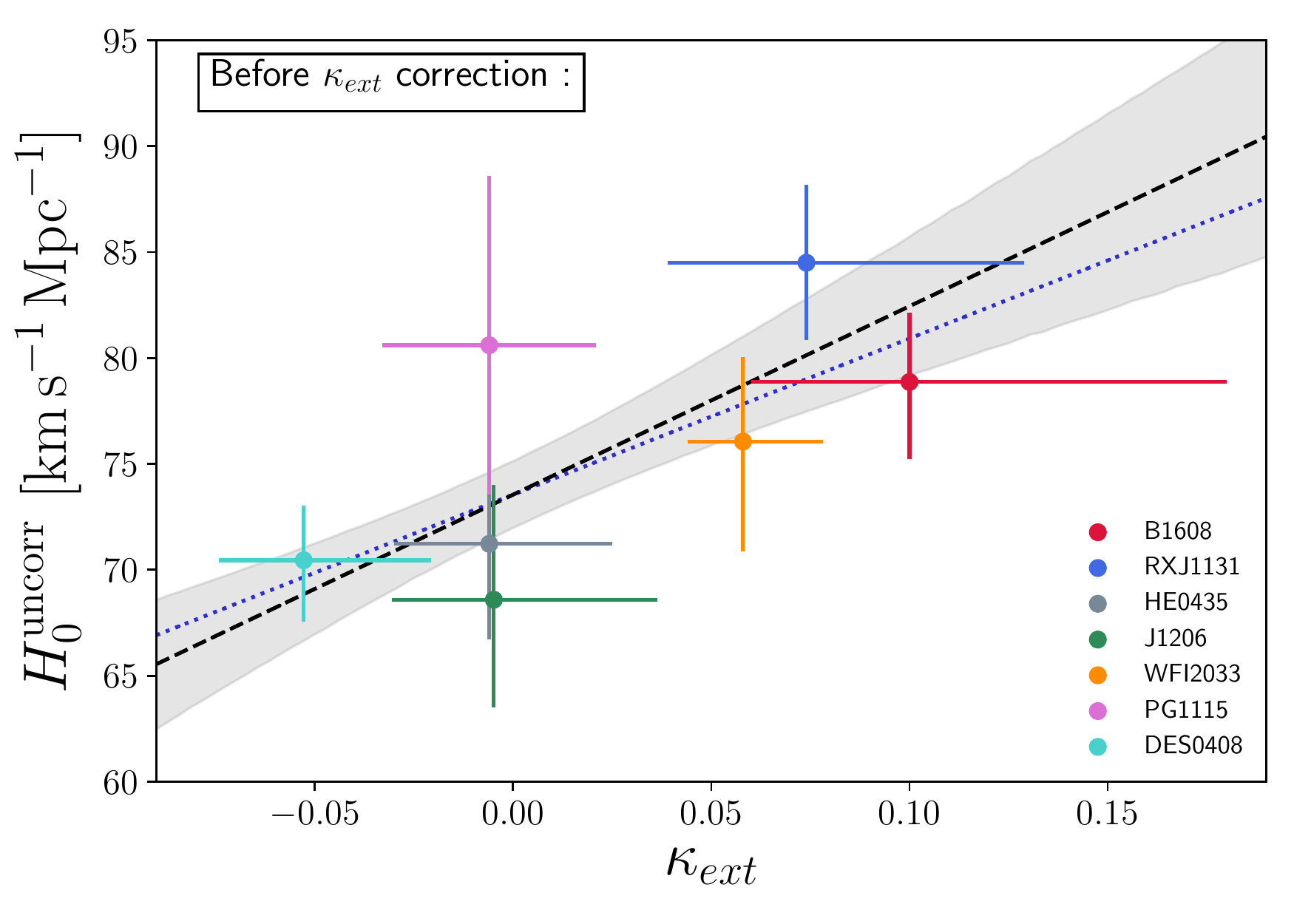}
    \includegraphics[width=\textwidth]{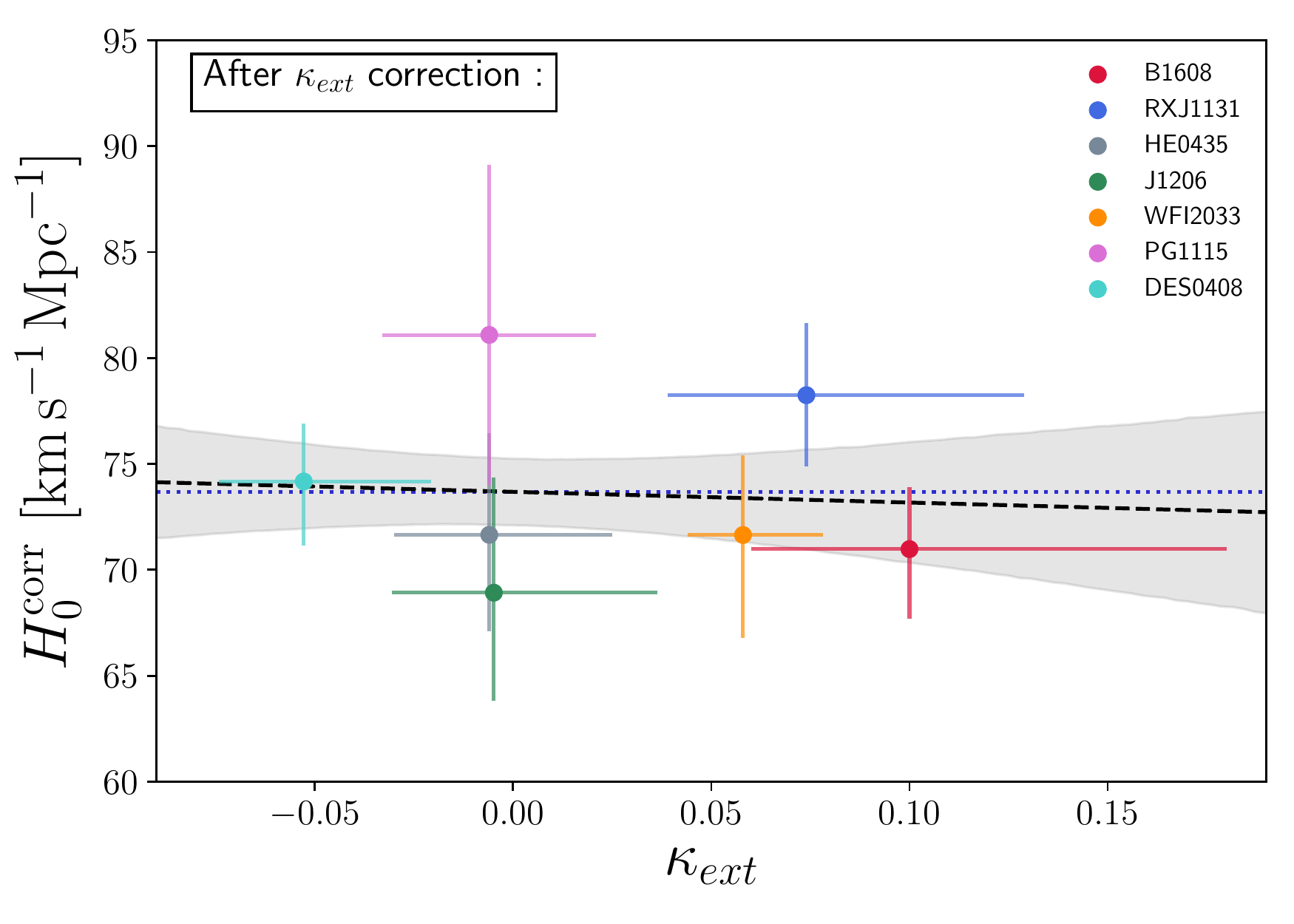}
    \end{minipage} 
    \caption{Measured Hubble constant, before (upper panel) and after (lower panel) correction for the mass along the line of sight as a function of the estimated external convergence. $\Hc^\mathrm{uncorr}$ and $\Hc^\mathrm{corr}$ are related according to Equation (\ref{eq:H0_externalk}). The dashed black lines show the best linear fit, and the shaded gray envelopes correspond to the 1$\sigma$ uncertainties. The dotted blue lines represent the  relation expected from the theory between $\Hc^\mathrm{uncorr}$, $\Hc^\mathrm{corr}$ and \kext.} 
    \label{fig:h0vskext}
\end{figure}

%%%%%%%%%%%%%%%%%%%
\subsection{Dependency on the external convergence and lens redshift}
\label{sec:kappa_ext}
%%%%%%%%%%%%%%%%%%%

In the previous sections, the focus is on how the lens velocity dispersion influences \Hc measurements. But there is also an external  contribution of all objects along the line of sight to the main lensing potential. This external convergence, \kext, is estimated in all TDCOSMO systems from galaxy counts, in combination with spectroscopy for obtaining redshifts for galaxies and quantifying coherent structures (e.g., groups and clusters). \cite{Tihhonova2018} showed that this measurement is compatible with the constraints obtained on \kext with weak lensing. \kext is directly related to the time-delay distance \Ddt , as shown in Equation (\ref{eq:MSD}). Similarly, the effect of the external convergence on the inferred \Hc can be written as : 
\begin{equation}
    \label{eq:H0_externalk}
    \Hc^\mathrm{uncorr} = \frac{\Hc^\mathrm{corr}}{(1 - \kext)}, 
\end{equation}
where $\Hc^\mathrm{uncorr}$ ($\Hc^\mathrm{corr}$) is the value of \Hc before (after) correction from \kext. 
The effect of this external MST can be mitigated by directly inferring \kext. To test the presence of residual external Mass-Sheet Degeneracy (MSD) not entirely removed by the measurement of \kext, we investigate the presence of correlation between the estimated \kext and the inferred \Hc value for the seven lenses of the TDCOSMO sample. The top panel of Fig.~\ref{fig:h0vskext} shows the relation between the \Hc measurements before correction for the mass along the line of sight $\Hc^\mathrm{uncorr}$, and the estimated convergence. A trend is visible between these two quantities indicating that the measurement is indeed sensitive to the lens environment. If no correction is applied, the lenses located in over-dense regions (positive \kext) tend to have a higher $\Hc^\mathrm{uncorr}$ than lenses in under-dense regions (negative \kext). We fit a linear model to the uncorrected data, and measure a slope of $a^\mathrm{uncorr} =88.9 \pm 29.1$\ksmpc, well compatible with the expected slope of $a^\mathrm{uncorr} = \Hc^\mathrm{corr} = 73.7$ \ksmpc. Both the uncorrected and corrected data are well fitted by our linear model, with a reduced $\chi^2$ of 0.61 and 0.95 respectively.

As shown on the bottom panel of Fig.~\ref{fig:h0vskext}, this trend disappears when correcting for the external convergence and there is no evidence for residual correlation between $\Hc^\mathrm{corr}$ and $\kext$. In fact, the best-fit slope coefficient in this case is $a^\mathrm{corr}=-5.1 \pm 23.7$\ksmpc, consistent with no correlation. This is an indication that the external convergence correction makes the trend disappear, which is what would be expected if our correction were accurately accounting for \kext. The present data set shows no evidence of residual systematic bias involving the LOS mass density.

\begin{figure}
    \centering
    \includegraphics[width=0.49\textwidth]{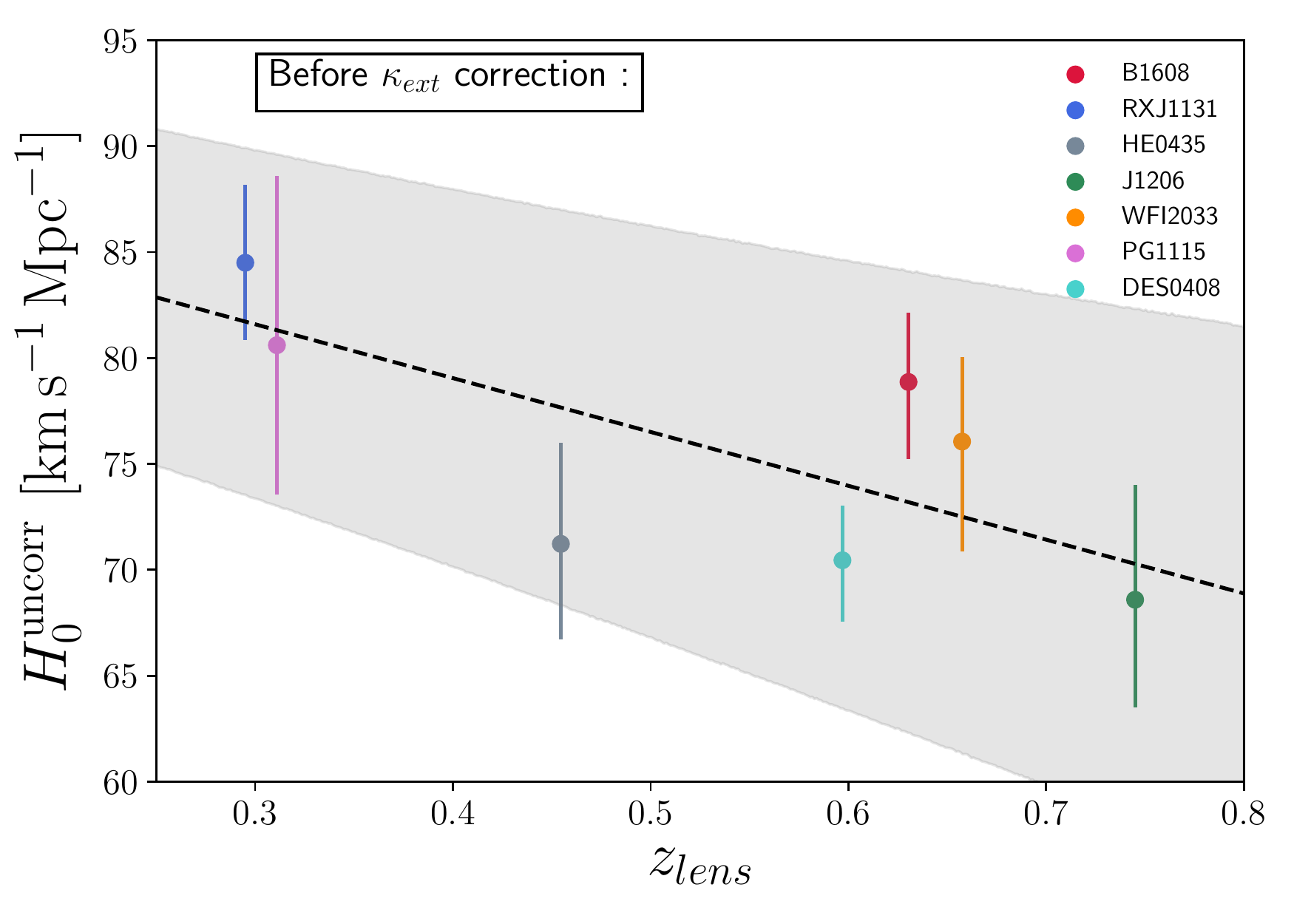}
    \includegraphics[width=0.49\textwidth]{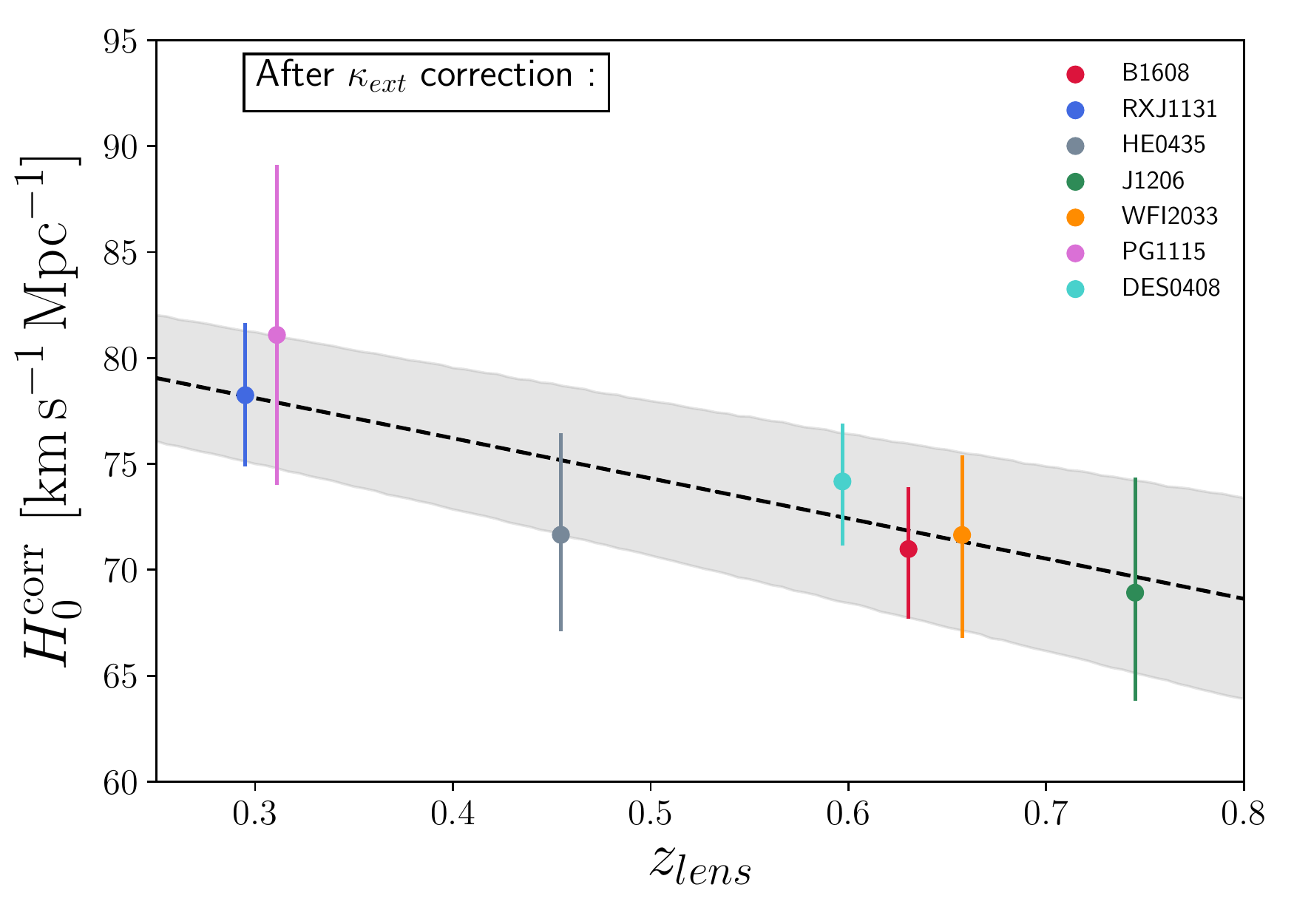}
    \caption{
     \Hc constraints for the TDCOSMO lenses as a function of lens redshift before (top) and after (bottom) correction for the external convergence. The best linear fits and their 1$\sigma$ envelopes are shown in shaded gray. The tentative ($1.7\sigma$ significance) trend is not introduced by the LOS contribution as it is still visible before correcting for the external convergence.}
    \label{fig:h0_kext_vs_zl}
\end{figure}{}

As first mentioned by \cite{Wong2019}, the H0LiCOW collaboration reported the presence of a possible trend between the lens redshift and the inferred $\Hc^\mathrm{corr}$ value at low statistical significance level ($\sim 1.9 \sigma$). When adding \DESzerofour to the six H0LiCOW lenses, the significance of the trend is slightly reduced to $\sim 1.7 \sigma$. We note that, having tested multiple correlations, it might be expected to find one at marginal significance, as a result of the "look elsewhere effect". This trend is still present before correction for the external convergence as shown on Fig.~\ref{fig:h0_kext_vs_zl}. The data are well-fitted by a linear model both before and after the LOS correction with a reduced $\chi^2$ of 1.52 and 0.24. The significance level of this correlation before LOS correction is still on the order of $\sim 2 \sigma$. Hence, there is no direct indication that the trend is due to unaccounted systematics in \kext. 

%%%%%%%%%%%%%%
\section{Impact of the choice of families of mass model}
\label{sec:6}
%%%%%%%%%%%%%%
\begin{figure*}[t!]
    \centering
    \includegraphics[width=0.49\textwidth]{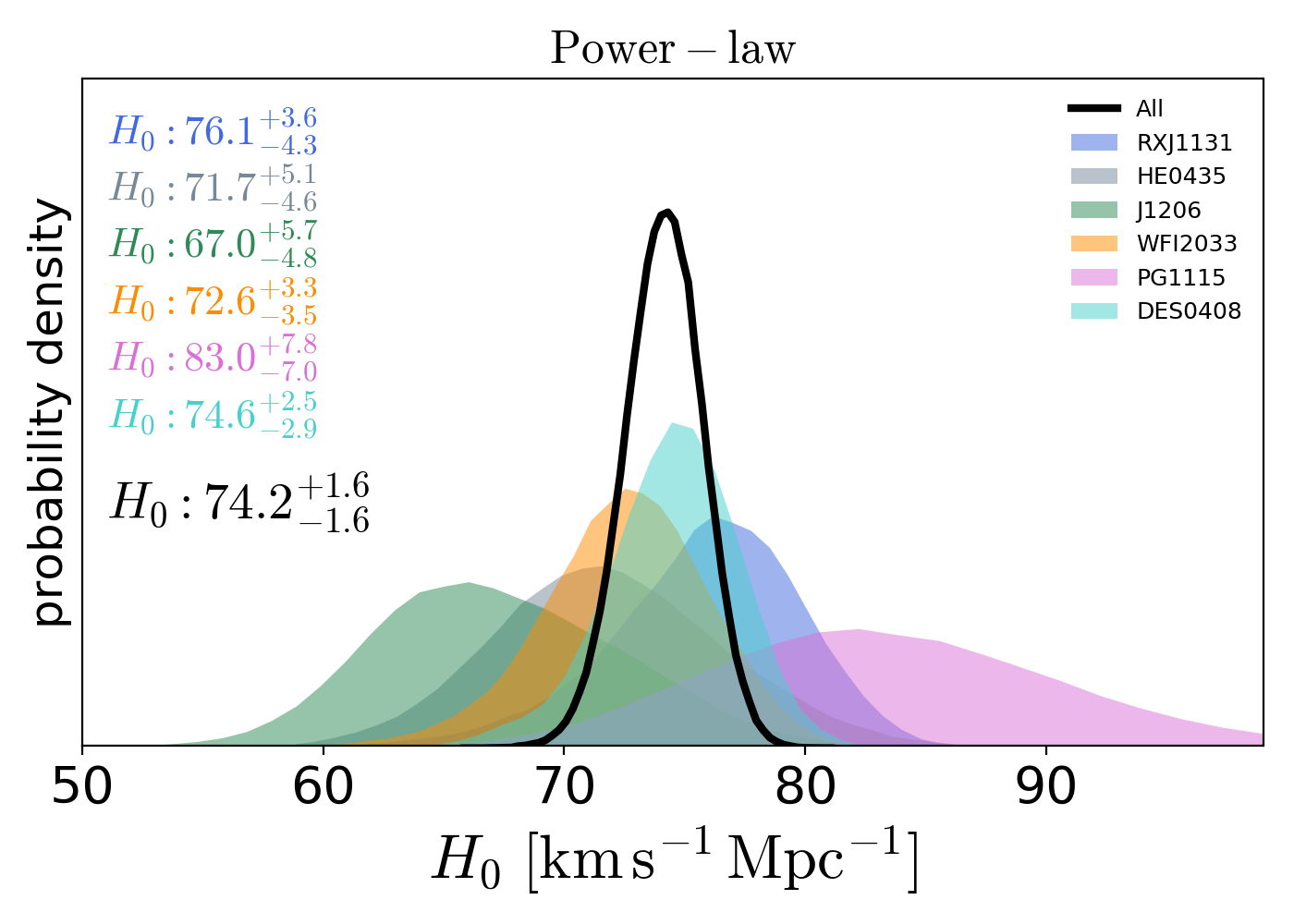}
     \includegraphics[width=0.49\textwidth]{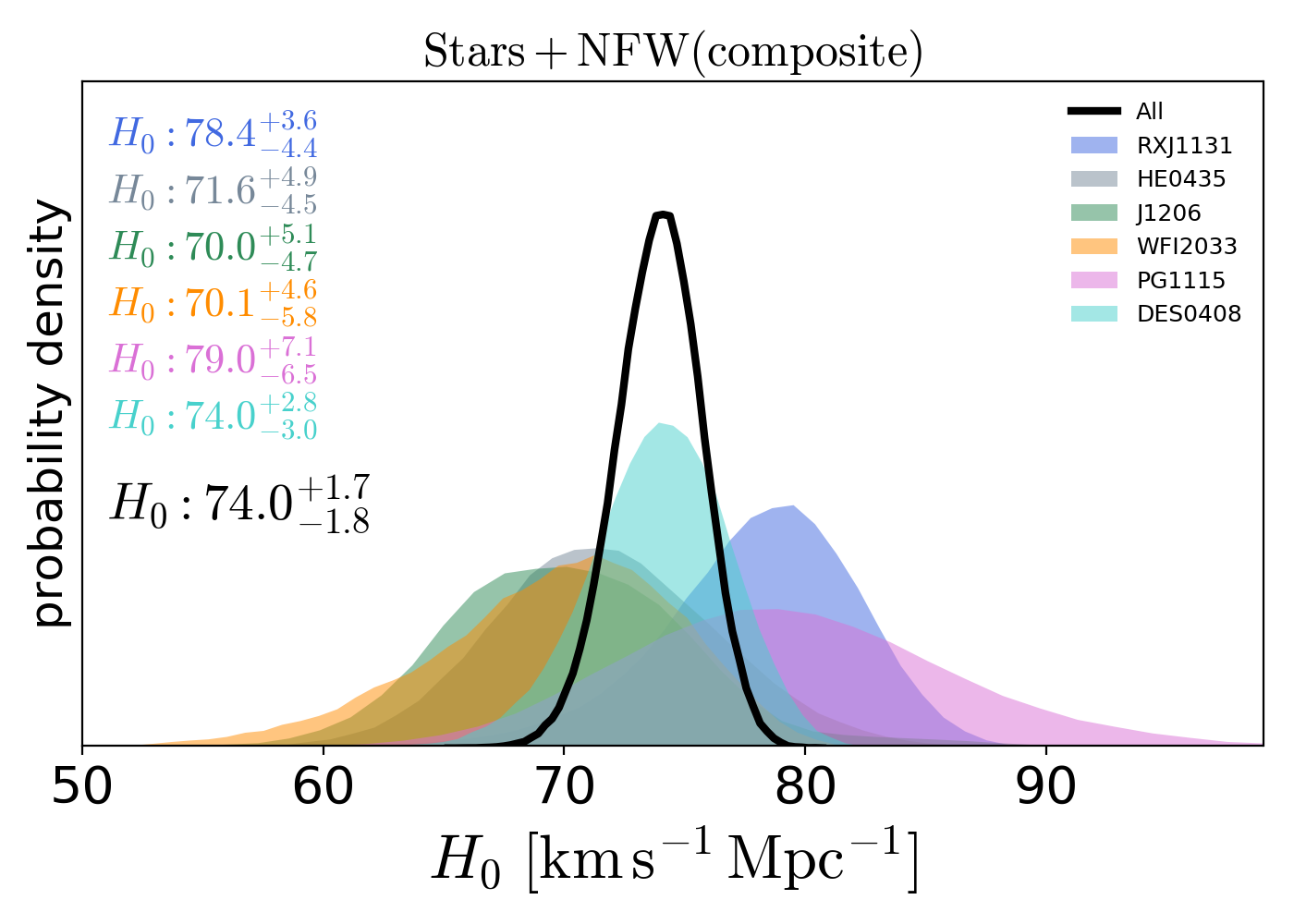}
    \caption{Marginalized $H_0$ posteriors for power-law (left panel) and composite models (right panel). The cosmological inferences are for a flat \lcdm cosmology with uniform priors. The posterior probability distributions for each individual system are shown with shaded color curves and the combined constraint from the six systems corresponds to the solid black curve. The legend indicates the median, 16$^{th}$ and 84$^{th}$ percentiles of the \Hc distributions.}
    \label{fig:H0_PL_composite}
\end{figure*}

In this section we quantify how much the inference on \Hc depends on the choice of the mass density profile adopted for the lens modeling. We first use the six systems for which both power-law and composite mass models have been performed and compare the results. We show that even though the two model families have sufficient flexibility to produce a broad range of profile shapes, in practice when applied to real elliptical massive galaxies, they form mass density profiles close to a simple power law. As we see below, this is likely due to the "bulge-halo" conspiracy \citep{TreuKoopmans2004,DuttonTreu2014}.

Then, we carry out end-to-end simulations in order to quantify the flexibility of our models and how the data actually allows us to constrain them. Meeting this goal requires the simulated properties of lenses to be close enough to those of real galaxies. About 90\% of galaxy-scale lenses are early-type galaxies \citep{Auger2009}, which satisfy very tight correlations between their observable properties \citep{Auger2010}. This indicates a high degree of regularity in the relative distribution of dark and luminous matter, often referred as the ``bulge-halo conspiracy''. This bulge-halo conspiracy results in the total mass density profile of lenses being very close to a singular isothermal ellipsoid \citep[e.g.,][]{Koopmans2006, Koopmans2009, VandeVen2009, Cappellari2016}, even out to large radii \citep{Gavazzi2007, Lagattuta2010}.

Importantly, the simulations we use all consider spatially extended lensing information, spanning a large range in radial extension. This radial extent must provide sufficient leverage to inform us about any possible departures from a simple power law within the actual range of observables. A goodness-of-fit criterion is then used to verify that the model adopted is indeed a good description of the data. Models that are exclusively based on the positions of two or four multiple quasar images, rather than the full surface brightness distribution of its spatially-extended host galaxy, cannot provide an accurate account of the uncertainties from surface brightness modeling. Therefore, models based on two or four image positions cannot satisfy the above goodness-of-fit requirement, even if they include time delays and stellar-velocity-dispersion measurements. In the following, we describe our set of simulated lenses in Section~\ref{ssec:sim}, present the results in Section~\ref{ssec:results} and discuss our findings in Section~\ref{ssec:disc}. 

%%%%%%%%%%%%%%%%%%
\subsection{\Hc inference per model family}
\label{ssec:powervscomposite}
%%%%%%%%%%%%%%%%%%

The TDCOSMO collaboration uses both composite and power-law models in their analysis, except for \Bsixteen (see Section~\ref{sec:current_model_familly}). Apart from this exception, the published estimates of \Hc correspond to the marginalization over the two model families as a way to account for modeling uncertainties \citep{Wong2019, Shajib2019}. The sample size of real lenses is now sufficiently large to infer \Hc by model family and to test whether this choice makes a difference at the 2\% precision level of the statistical uncertainty. This is illustrated in Fig.~\ref{fig:H0_PL_composite}, where the priors on the cosmological parameters are the same as adopted by \citet{Wong2019}: $H_0 \in [0, 150]$\ksmpc, $\Omega_{\rm m} \in [0.05, 0.5]$ and $\Omega_{\rm m} = 1  - \Omega_{\Lambda}$. 

The \Hc values vary with the model family for individual objects, and this testifies to the flexibility of the families of models. However, the choice of model family changes the combined value by much less than the estimated statistical uncertainty. Quantifying these statements, the combined value from the six lenses is $\Hc = 74.2^{+1.6}_{-1.6}$ \ksmpc when we use exclusively power-law models and $\Hc = 74.0^{+1.7}_{-1.8}$ \ksmpc when we use only composite model. This corresponds only to a 0.2\% difference. Individual objects can have larger differences between power-law and composite models than the combined estimate, but the two posterior probability distributions always remain compatible. The largest differences are found for \PGeleven (5\%) and \Jtwelve (4\%), which still have the two distributions compatible at the $\sim 0.6\sigma$ level. 

Last but not least, there is no indication in the current sample of six lenses that one given family of models systematically gives a lower or higher \Hc value. For example, \WFItwenty has a higher \Hc value when modeled with a power law rather than a composite, while the opposite behavior is found for \Jtwelve; and other such examples can be easily found in Fig.~\ref{fig:H0_PL_composite}.  

In conclusion, even though our two families of models are flexible enough to produce a broad range of \Hc values, in practice they do not. In the following, we investigate with simulated lens systems the reasons why composite and power-law models provide comparable estimates of \Hc in spite of allowing for flexibility. We also investigate  under which circumstances gravitational lenses can be modeled with both composite and power-law models and still yield the same \Hc.  

%%%%%%%%%%%%
\subsection{Simulations}
\label{ssec:sim}
%%%%%%%%%%%%
\begin{figure*}[h!]
    \centering
    \includegraphics[width=1\textwidth]{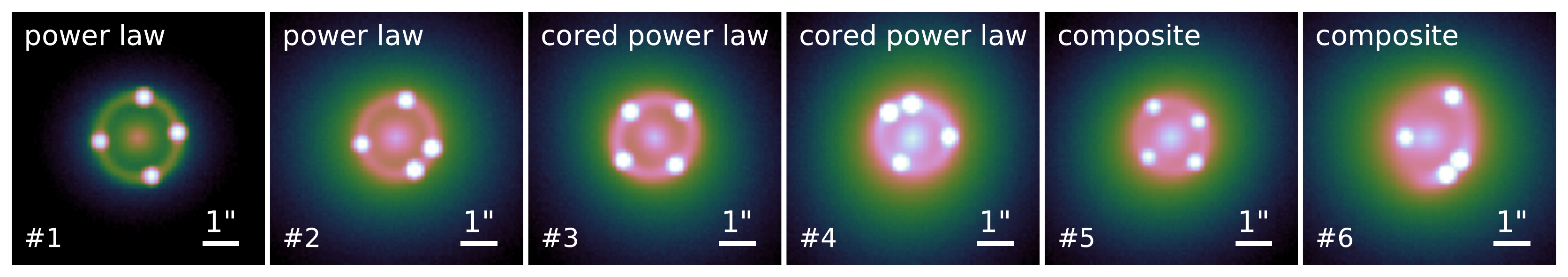}
    \caption{Sample of simulated lenses: three pairs are generated from power-law, cored power-law, and composite lens models. The color scale is logarithmic and is the same for all images. Identifiers associated to each lens are also indicated. Refer to Section~\ref{ssec:sim} for a description of these simulations. Model \#6, although composite, is chosen so that the total mass profile resembles a power law in the region of the Einstein radius.}
    \label{fig:mock_lenses}
\end{figure*}
%[Xuheng rewrite:]
We generate six mock lens systems chosen to illustrate the range of possible outcomes, labeled by IDs \#1 through \#6. We describe the process of the simulations in this section. In addition to the power-law and composite models typically used by TDCOSMO we also include cored power laws to explore the effects of adding extra flexibility to the models.

The simulated HST images are produced using the pipeline described by \citet{Ding2017a, Ding2018}. The image frame size is chosen to be $99\times99$ pixels, with a pixel scale of $0\farcs08$ to mimic the realistic HST WFC3/F160W drizzled resolution. Mass profile parameters are chosen such that the Einstein radius is roughly at the scale of $1''$ as typical for galaxy-scale lenses. The noise in each pixel is composed of the Gaussian background noise and the Poisson noise. For Gaussian background noise, we assume an rms of 0.003, which is directly measured from empty regions in the real data; the Poisson noise is added, based on a total exposure time of $2400$~s. For computational speed, the PSF is assumed as a Gaussian kernel with FWHM=$0\farcs25$.

Three mass models, including power-law (ID \#1, \#2), cored power-law (ID \#3, \#4), and composite (ID \#5, \#6) mass density profiles, are adopted to generate the six mock systems. All of the systems are elliptical in projection in order to allow for quad-like configurations by construction. For each family of mass distribution, we generate two mock lensed systems, one with the source lying close to a fold of a caustic ("fold" configuration) and one with the source lying close to the lens-optical axis ("cross-like" configuration). The "cross" represents a worst case scenario because the radial ranges and the differences in the time delays are limited by symmetry. The simulated lens systems are shown in Fig.~\ref{fig:mock_lenses}.

For the composite model, the total mass consists of a baryonic elliptical Hernquist profile \citep{Hernquist1990}, and a dark matter elliptical NFW profile \cite[see Eq.~\ref{eq:nfw} and][]{Navarro1997}. The baryonic part is linked to the lens surface brightness through a constant mass-to-light ratio. While we use the same axis ratios for the baryonic and dark matter components, we allow for slight offsets in their position angles; the total projected mass profile is therefore not elliptical. We note that the system (ID \#6) is chosen to describe a scenario similar to realistic galaxies, in which luminous and dark matter conspire to produce a total mass model very close to a power-law profile. This is consistent with the findings of the H0LiCOW, SHARP, and STRIDES collaborations so far \citep{Suyu2014, Wong2017, Birrer2019, Chen2019, Rusu2019, Shajib2019}. Other cored power-law and composite systems (ID  \#3 -- \#5) are designed on purpose to depart significantly from a single power law in order to test the effect on \Hc and investigate whether the information contained in the data can capture this discrepancy. For all the lenses, the deflector surface brightness is simulated as an elliptical Hernquist profile. The ellipticity of the simulated lens galaxy corresponds to an axis ratios of $q\sim 0.9\pm0.01$. We use an elliptical S\'ersic profile \citep{Sersic1963} to simulate the extended part of the source light, which is sufficient for our purpose. Lensed quasar images are modeled as point spread functions centered on the images of the host galaxy.

The simulated time delays are calculated within a fiducial flat \lcdm cosmology with $\Omega_{\mathrm{m}}=0.27$, and $\Omega_\Lambda=0.73$, and Hubble constant $H_0^\mathrm{fiducial}=70.7$ \ksmpc, which was chosen randomly. For the time-delay uncertainties, we assume an unbiased random error with rms level set as the largest value between $\Delta t\times1\%$ and $0.25$ days. The uncertainties on the time delays are chosen to be smaller than current uncertainties of real data in order to focus mainly on the modeling uncertainties. 

Since the tests in this section focus on the mass reconstruction of the main deflector, we do not include in the simulations the effects of the galaxies along the line of sight, which are treated separately in real data. Likewise, we simulate and model the velocity dispersion using spherical Jeans equations following \cite{Suyu2010} and \cite{Birrer2019}, and assume an anisotropy radius equal to the lens half-light radius. This is a simplification of the stellar kinematics treatment with respect to the analysis of real systems where TDCOSMO marginalizes over the unknown anisotropy. In this exercise where we aim to illustrate the constraining power of the images while saving computing time, we do not use the LOS velocity dispersion as a direct constraint in the modeling but rather only calculate the modeled values to make the comparison with measured values. The relevant key properties of the six simulated lenses are summarized in Table~\ref{app:mock_params}.

\subsection{Results}
\label{ssec:results}
\begin{table*}[htbp!] 
\renewcommand{\arraystretch}{1.2}
\centering
\small
\begin{tabular}{l|ccc}
& Model: power law             & Model: cored power law       & Model: composite         \\ \hline \hline 
Truth: power law (\#1)& BIC = 10220& BIC = 10230\\ 
	& - & $\Delta$ BIC = 10\\ 
	($\sigma_v = 308$\,\ks) & $\chi^2 = 1.02$ & $\chi^2 = 1.02$\\
	& $H_0 = 71.9 ^{+2.1}_{-2.3}$\,\ksmpc & $H_0 = 70.9 ^{+2.2}_{-2.0}$ \,\ksmpc & - \\ 
	& Tension = 0.5 $\sigma$ & Tension = 0.1 $\sigma$ \\ 
	& $\sigma_v = 310.1 ^{+1.4}_{-1.5}$\,\ks& $\sigma_v = 304.3 ^{+3.9}_{-3.4}$\,\ks\\ \\ 
Truth: power law (\#2)& BIC = 9786& BIC = 9797\\ 
	& - & $\Delta$ BIC = 11\\ 
	($\sigma_v = 297 $\,\ks) & $\chi^2 = 0.98$ & $\chi^2 = 0.98 $\\
	 & $H_0 = 72.6 ^{+1.8}_{-1.7}$\,\ksmpc & $H_0 = 72.2 ^{+2.0}_{-2.0}$\,\ksmpc & - \\ 
	& Tension = 1.1 $\sigma$ & Tension = 0.8 $\sigma$ \\ 
	& $\sigma_v = 298.1 ^{+1.1}_{-1.0}$\,\ks& $\sigma_v = 296.3 ^{+1.6}_{-1.6}$\,\ks \\ \hline 
Truth: cored power law (\#3)& BIC = 14544& BIC = 9776\\ 
	& $\Delta$ BIC = 4768& - \\ 
	($\sigma_v = 245 $\,\ks ) & $\chi^2 = 1.46$ & $\chi^2 = 0.98 $\\
	& $H_0 = 76.3 ^{+2.1}_{-2.0}$\,\ksmpc & $H_0 = 72.3 ^{+2.1}_{-2.2}$\,\ksmpc & - \\ 
	& Tension = 2.8 $\sigma$ & Tension = 0.7 $\sigma$ \\ 
	& $\sigma_v = 248.4 ^{+0.6}_{-0.7}$\,\ks & $\sigma_v = 245.3 ^{+1.2}_{-1.2}$\,\ks \\ \\ 
Truth: cored power law (\#4)& BIC = 18565& BIC = 9768\\ 
	& $\Delta$ BIC = 8797& - \\ 
	($\sigma_v = 216 $\,\ks ) & $\chi^2 = 1.86$ & $\chi^2 = 0.98 $\\
	& $H_0 = 78.2 ^{+1.9}_{-2.0}$\,\ksmpc & $H_0 = 71.8 ^{+1.5}_{-1.8}$\,\ksmpc & - \\ 
	& Tension = 3.9 $\sigma$ & Tension = 0.7 $\sigma$ \\ 
	& $\sigma_v = 219.0 ^{+0.6}_{-0.6}$\,\ks & $\sigma_v = 216.1 ^{+1.0}_{-1.2}$\,\ks \\ \hline 
Truth: composite (\#5)& BIC = 10042& BIC = 9703& BIC = 9608\\ 
	& $\Delta$ BIC = 434& $\Delta$ BIC = 95& - \\ 
	($\sigma_v = 253 $\,\ks ) & $\chi^2 = 1.00$ & $\chi^2 = 0.97 $ & $\chi^2 = 0.96 $\\
	 & $H_0 = 63.9 ^{+1.3}_{-1.1}$\,\ksmpc & $H_0 = 60.4 ^{+1.1}_{-1.2}$\,\ksmpc & $H_0 = 69.0 ^{+2.4}_{-2.7}$\,\ksmpc \\ 
	& Tension = 5.2 $\sigma$ & Tension = 9.1 $\sigma$ & Tension = 0.7 $\sigma$ \\ 
	& $\sigma_v = 259.5 ^{+0.6}_{-0.6}$\,\ks & $\sigma_v = 243.1 ^{+1.1}_{-0.9}$\,\ks & $\sigma_v = 255.7 ^{+1.6}_{-2.0}$\,\ks \\ \\ 
Truth: composite (\#6)& BIC = 14170& BIC = 10764& BIC = 9715\\ 
	& $\Delta$ BIC = 4455& $\Delta$ BIC = 1049& - \\ 
	($\sigma_v = 207 $\,\ks ) & $\chi^2 = 1.36$ & $\chi^2 = 1.04 $ & $\chi^2 = 0.97 $\\
	 & $H_0 = 69.8 ^{+1.1}_{-1.2}$\,\ksmpc & $H_0 = 70.0 ^{+1.2}_{-1.2}$\,\ksmpc & $H_0 = 72.4 ^{+1.9}_{-1.7}$\,\ksmpc \\ 
	& Tension = 0.8 $\sigma$ & Tension = 0.6 $\sigma$ & Tension = 1.0 $\sigma$ \\ 
	& $\sigma_v = 200.5 ^{+0.6}_{-0.8}$\,\ks & $\sigma_v = 200.7 ^{+0.9}_{-0.8}$\,\ks & $\sigma_v = 211.7 ^{+1.5}_{-1.2}$\,\ks \\ \\ 
\end{tabular}
\caption{\label{tab:test_modeling} BIC value, reduced $\chi^2$ of the image fit, measured \Hc, tension relative to the true value of $\Hc^\mathrm{fiducial}=70.7$ \ksmpc and integrated stellar velocity dispersion. The three columns of the table correspond to the family of mass model fitted on the six simulated lens systems. The $\Delta$BIC is computed relative to the best model for each lens.}
\end{table*}

The six mock lenses are modeled using the public strong lensing modeling package \LENSTRO\footnote{\url{https://github.com/sibirrer/lenstronomy}} \citep{Birrer2015, Birrer2018}, which was used for the latest analysis of the real systems \Jtwelve and \DESzerofour \citep{Birrer2019, Shajib2019}. The exact/known input PSF is used as the effect of PSF imperfections is not investigated in this work. The light profile of the lens and of the source are modeled as Hernquist and S\'ersic profiles respectively. We fit three types of analytical elliptical mass profiles to the simulated data, namely a power-law, a cored power-law and a composite profile. Specifically for the composite model, we emphasize that no strong prior is applied on the scale radius of the dark matter component. Instead, we use a noninformative uniform prior $r_s\sim\mathcal{U}\left(5'',\,40''\right)$, so that the dark matter component effectively has two degrees of freedom in the radial direction. The $99\times99$ pixels contained in the images and three independent time delays are used for the fit. We, however, mask a central region corresponding to three pixels (i.e., $0\farcs24$) since we do not want to form any central image which could lead to extra constraints on the lens model \citep[see also][]{Tagore2018, Mukherjee2018, Mukherjee2019}. The resulting fitted models are used to infer only \Hc ($\Omega_{\mathrm{m}}$ is kept fixed to 0.27) from the time-delay distance alone. The lens velocity dispersion is computed only for comparison but is not included in the \Hc inference, to highlight the information content of the images.

We use the Bayesian Information Criterion (BIC) to evaluate the quality of the fit. The BIC is defined by
\begin{equation}
\label{eq:BIC}
{\rm BIC} = k\times \mathrm{ln}(n) - 2\times \mathrm{ln}(\hat{L}), 
\end{equation}
where $k$ is the number of free parameters, $\hat{L}$ is the maximum likelihood of the model and $n$ is the number of data points. The likelihood used for the fit uses only the imaging and time-delay information so that $n$ corresponds to the number of nonmasked pixels in the image plus the three time delays. Our models have 25 free parameters for the power-laws, 26 for the cored power-laws and 29 for the composite models.

The recovered \Hc value, integrated LOS velocity dispersion within a square aperture of side $1''$ and the BIC values are given in Table~\ref{tab:test_modeling}. The corresponding image residuals of the lens modeling are shown in Table \ref{tab:test_residuals}. As expected, we recover the correct \Hc value within the 1-$\sigma$ errors of the posterior distribution when fitting the same mass model family as used in the simulation. This case corresponds to the diagonal of Tables~\ref{tab:test_modeling} and \ref{tab:test_residuals}. 

\begin{table*}[htbp!] 
\centering
\renewcommand{\arraystretch}{1.1}
\small
\begin{tabular}{l|m{3.5cm}m{3.5cm}m{3.5cm}}
& Model: power law             & Model: cored power law        & Model: composite         \\ \hline \hline 
Truth: power law (\#1) &\includegraphics[width=0.14\textwidth]{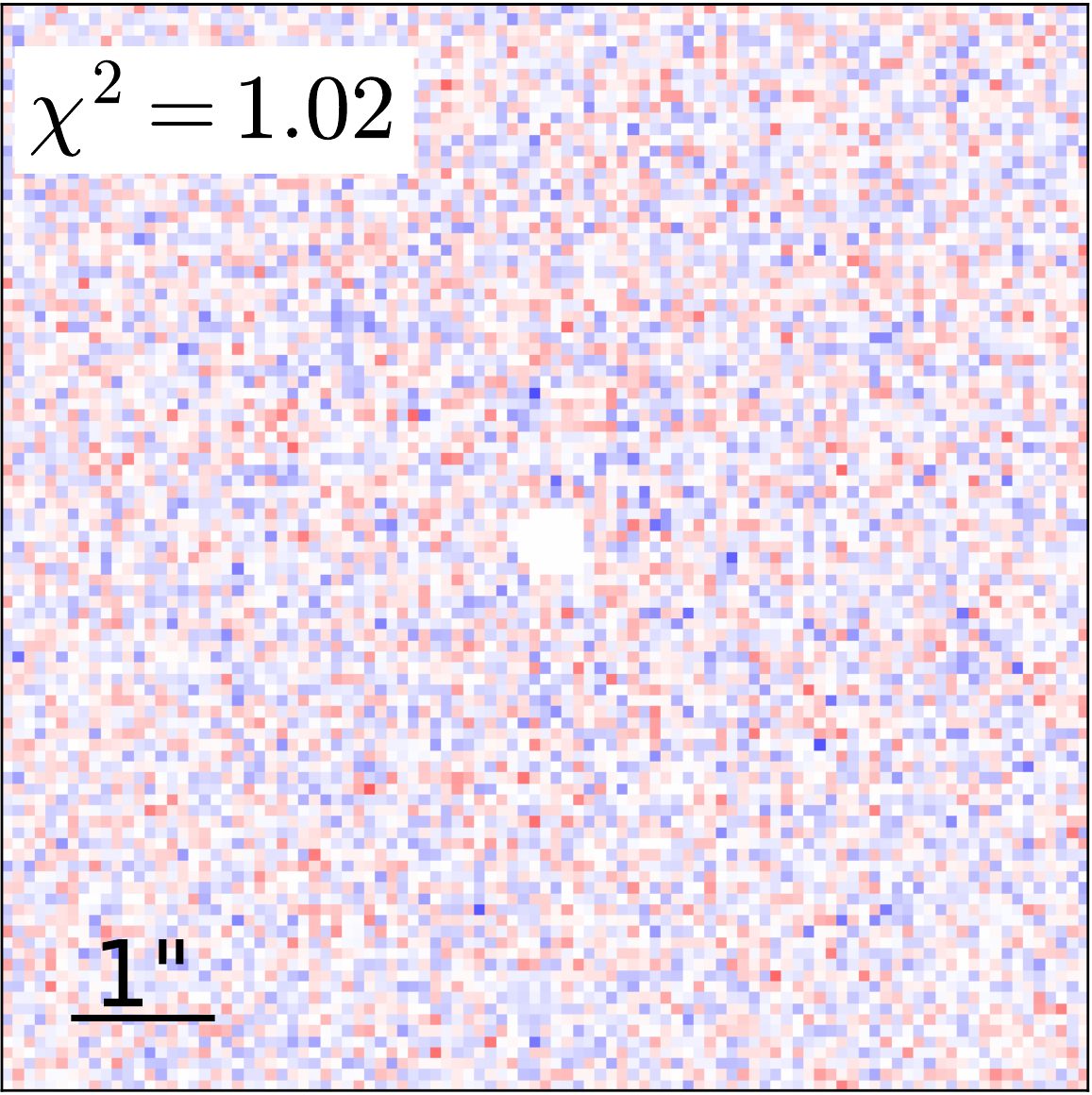} &\includegraphics[width=0.14\textwidth]{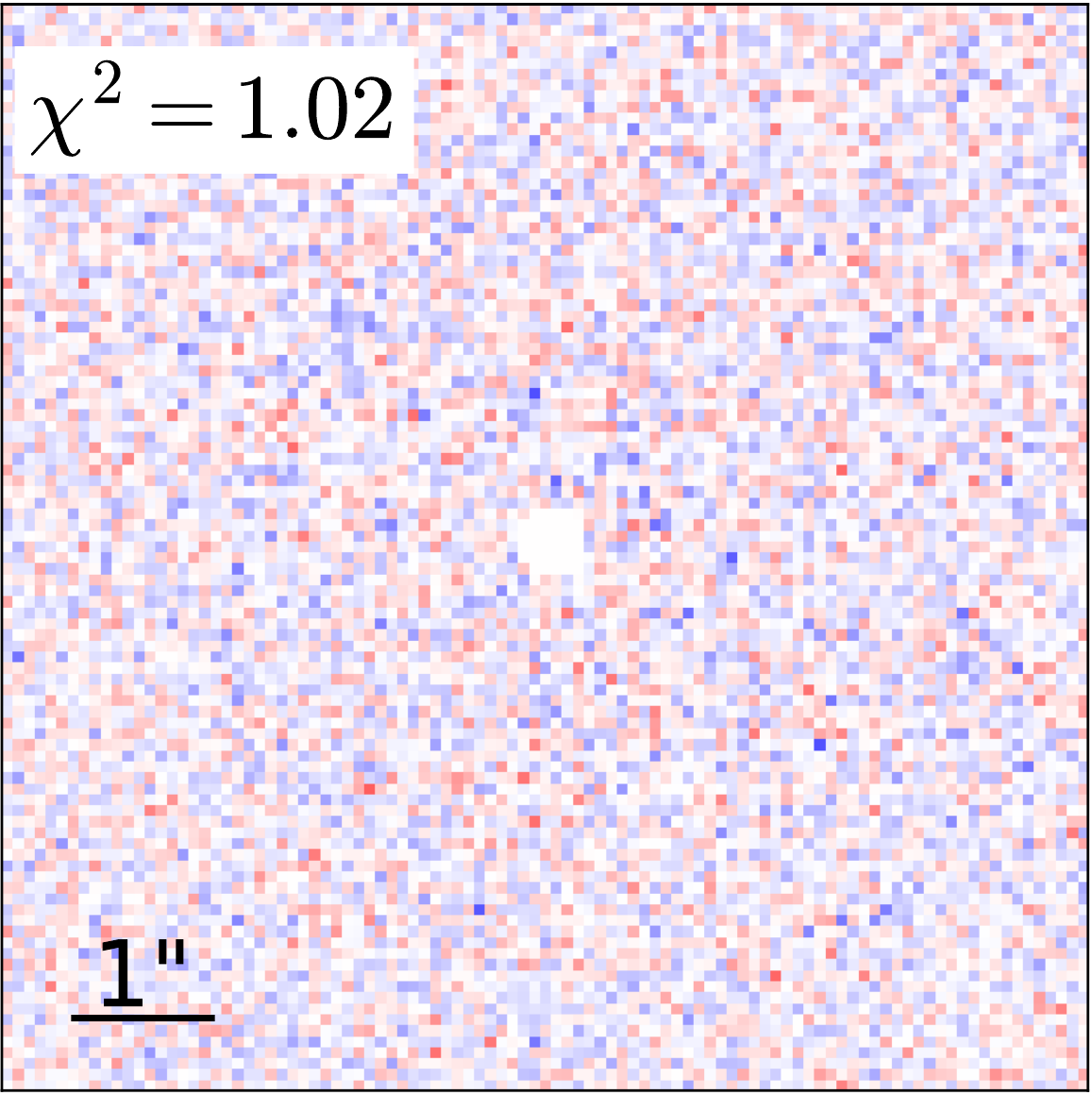} \\ 
Truth: power law (\#2) &\includegraphics[width=0.14\textwidth]{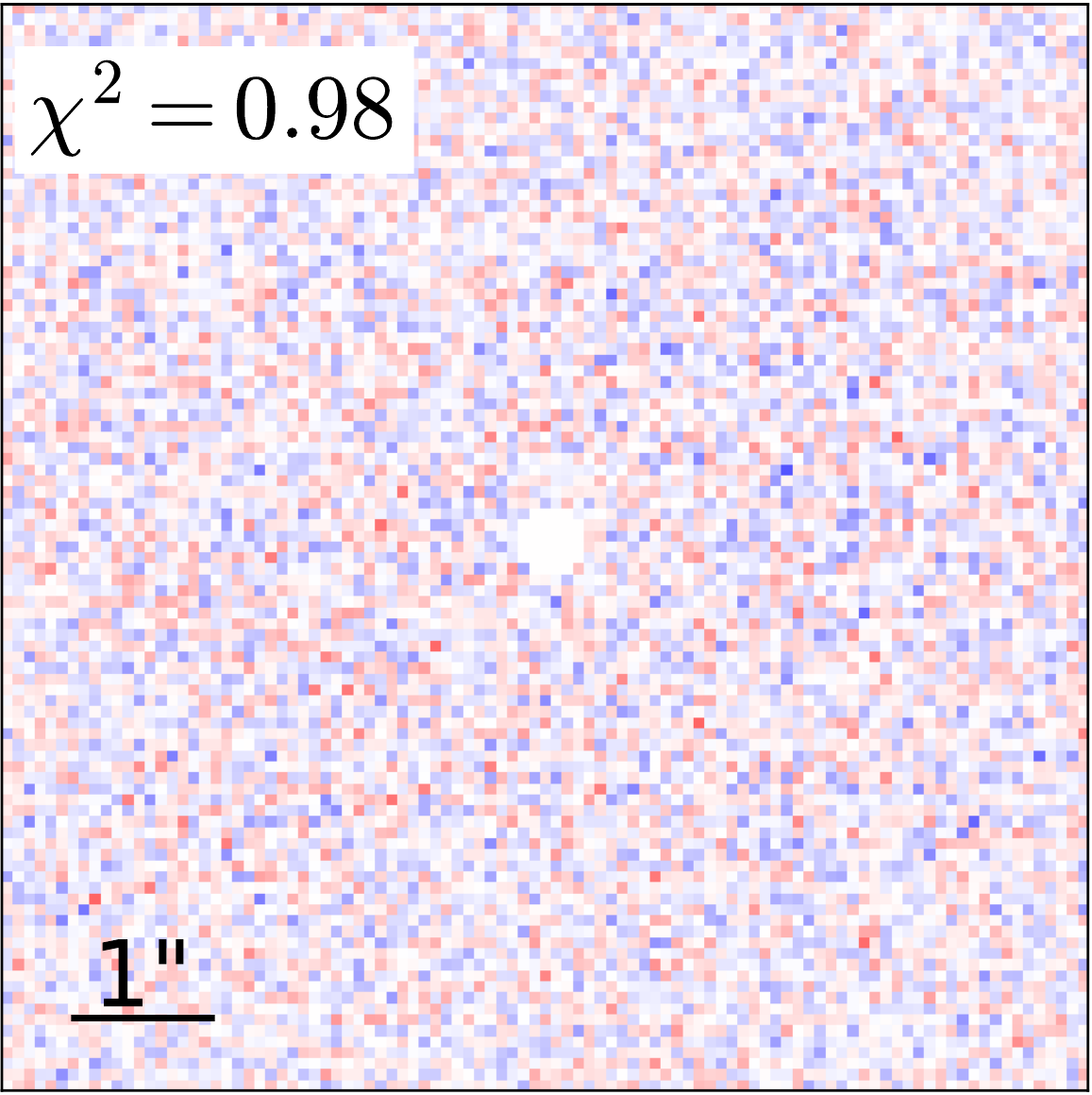} &\includegraphics[width=0.14\textwidth]{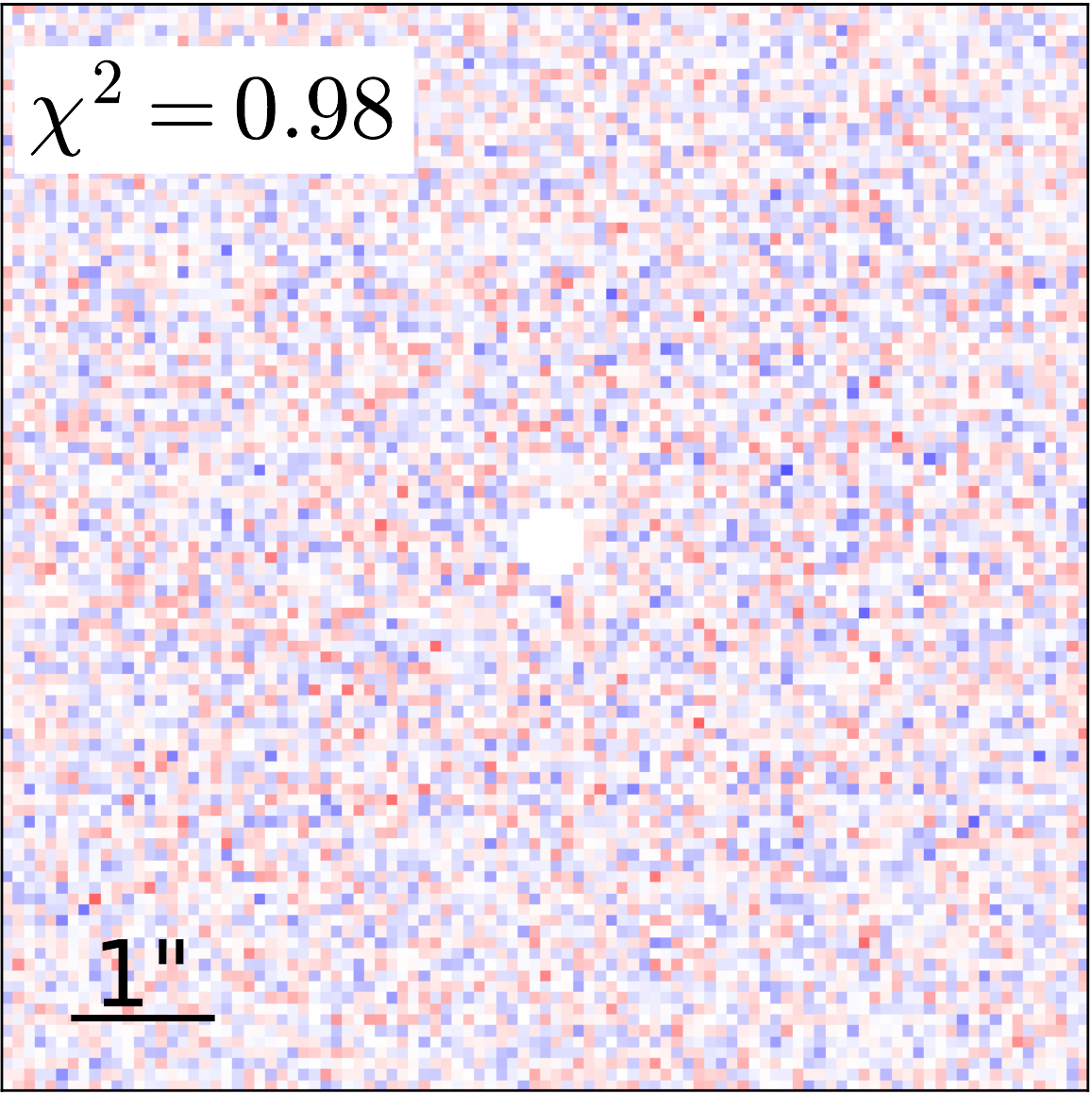} \\  \hline 
Truth: cored power law (\#3) &\includegraphics[width=0.14\textwidth]{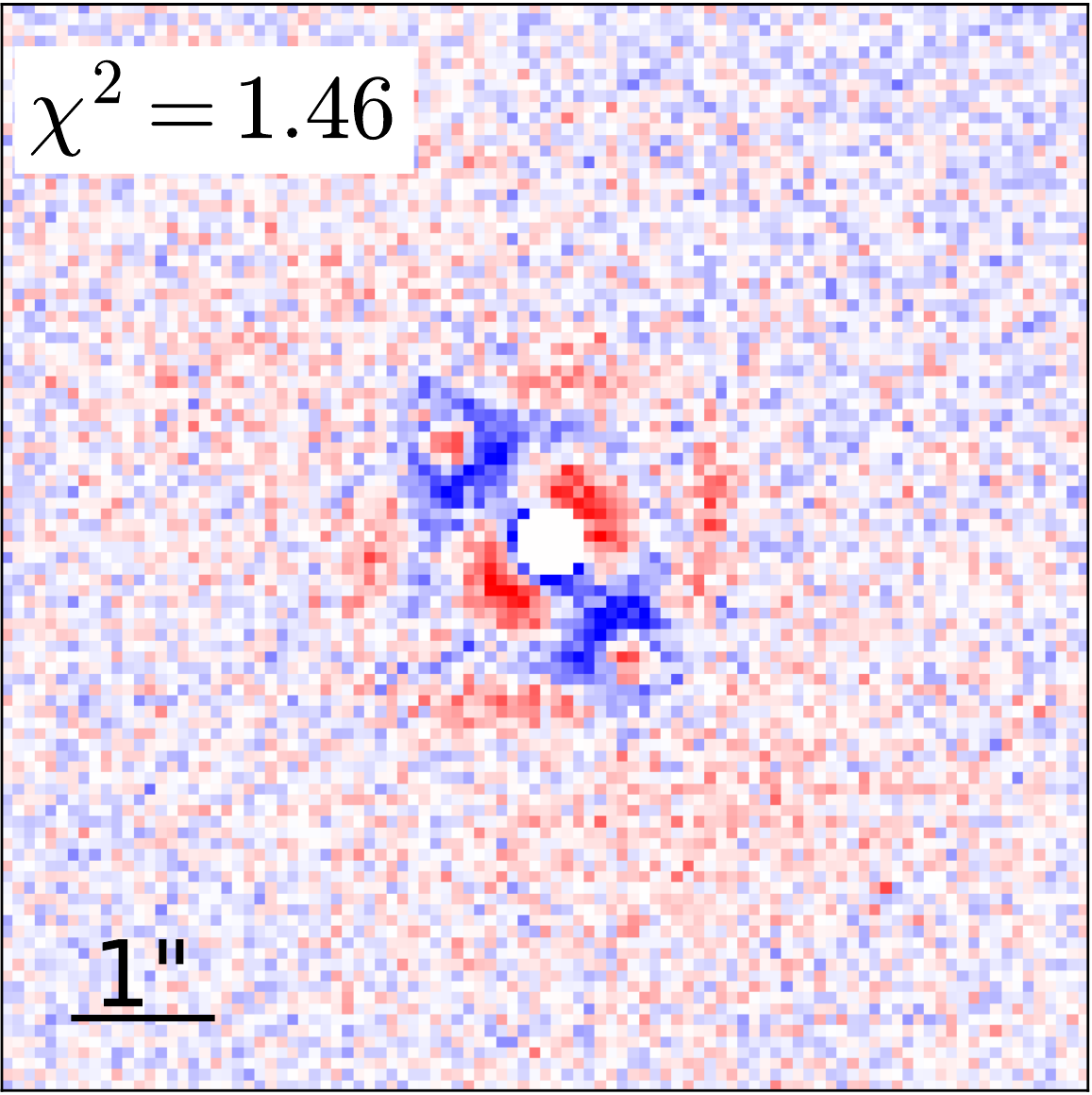} &\includegraphics[width=0.14\textwidth]{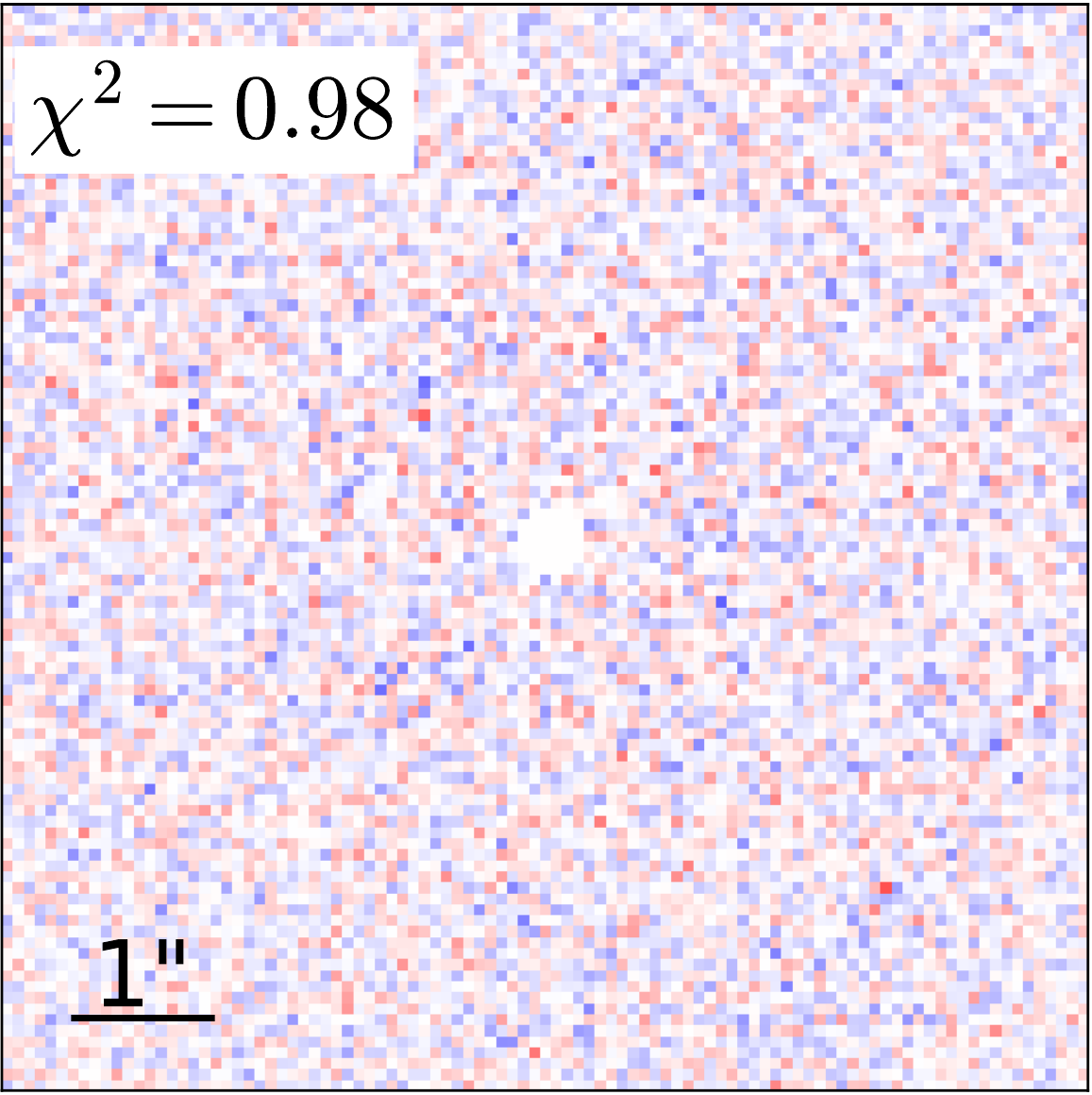} \\ 
Truth: cored power law (\#4) &\includegraphics[width=0.14\textwidth]{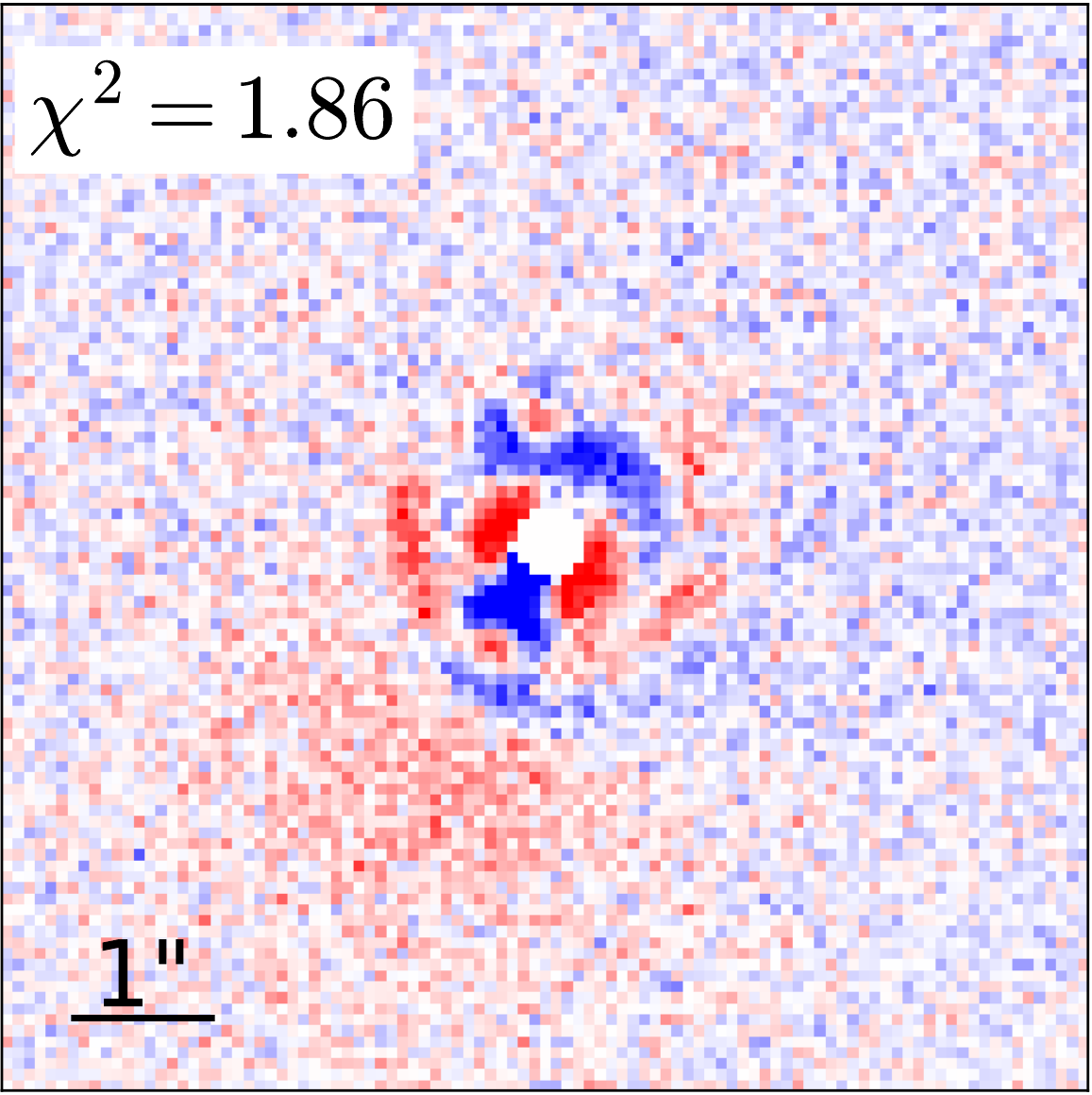} &\includegraphics[width=0.14\textwidth]{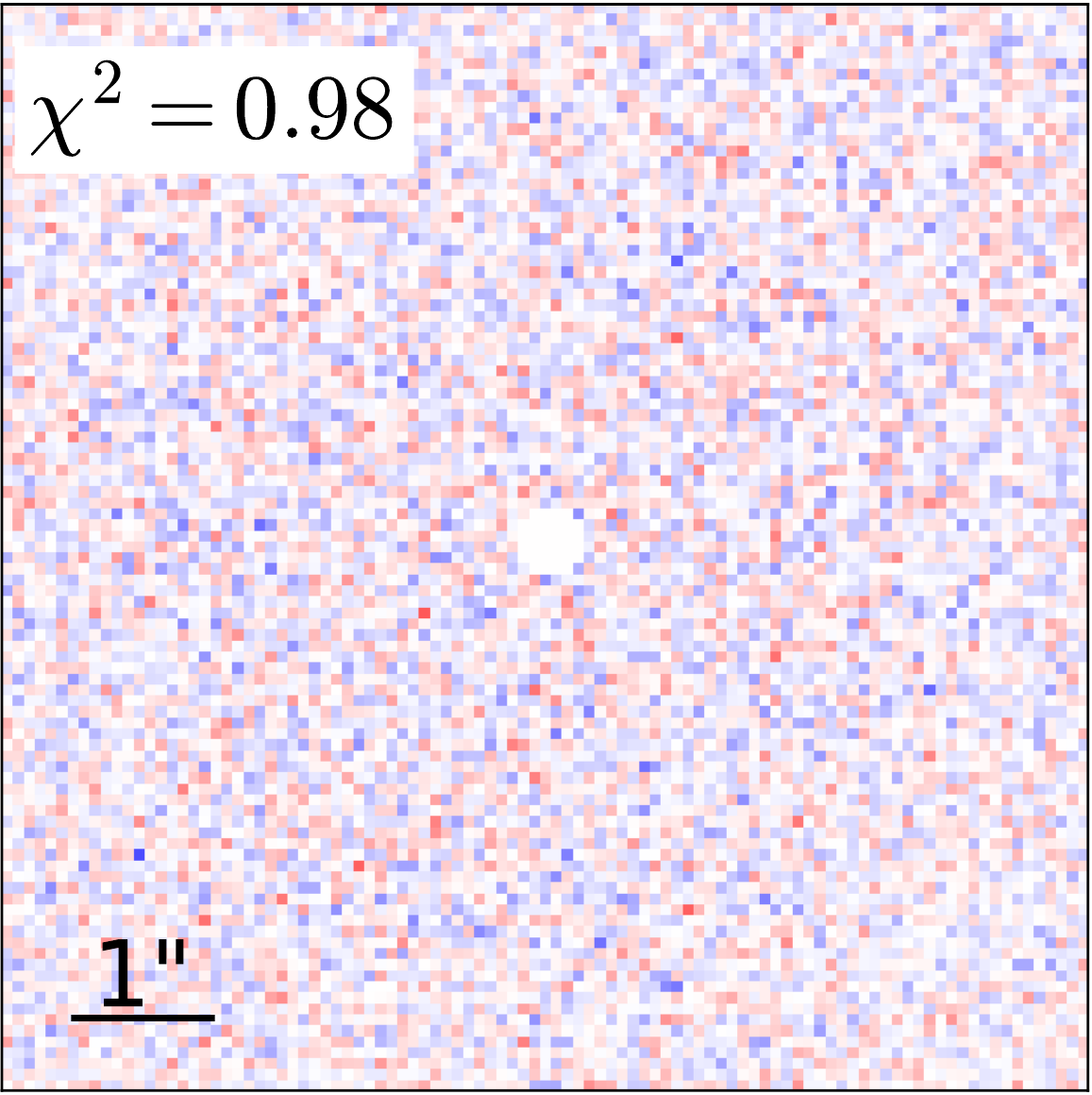} \\  \hline 
Truth: composite (\#5) &\includegraphics[width=0.14\textwidth]{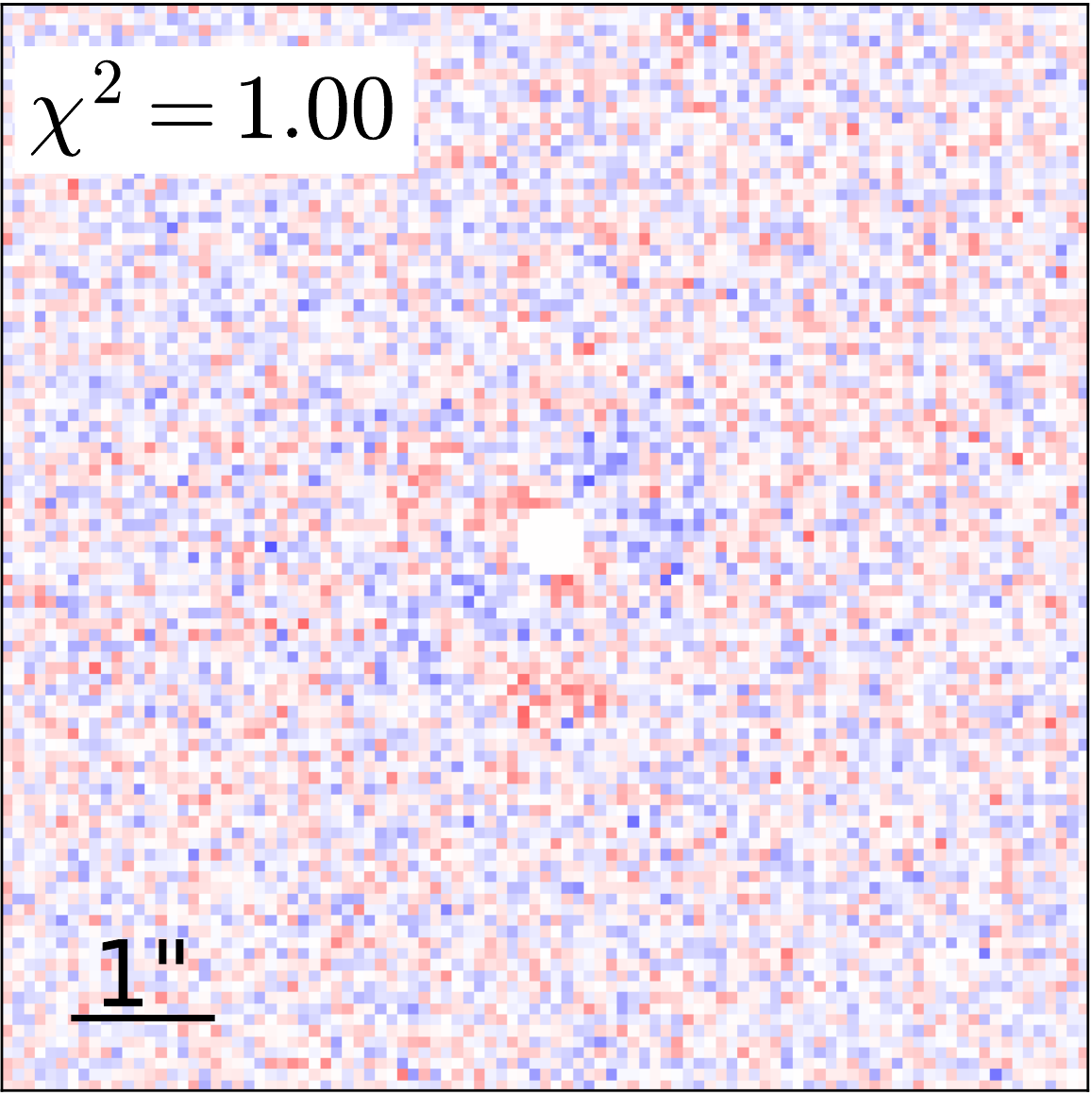} &\includegraphics[width=0.14\textwidth]{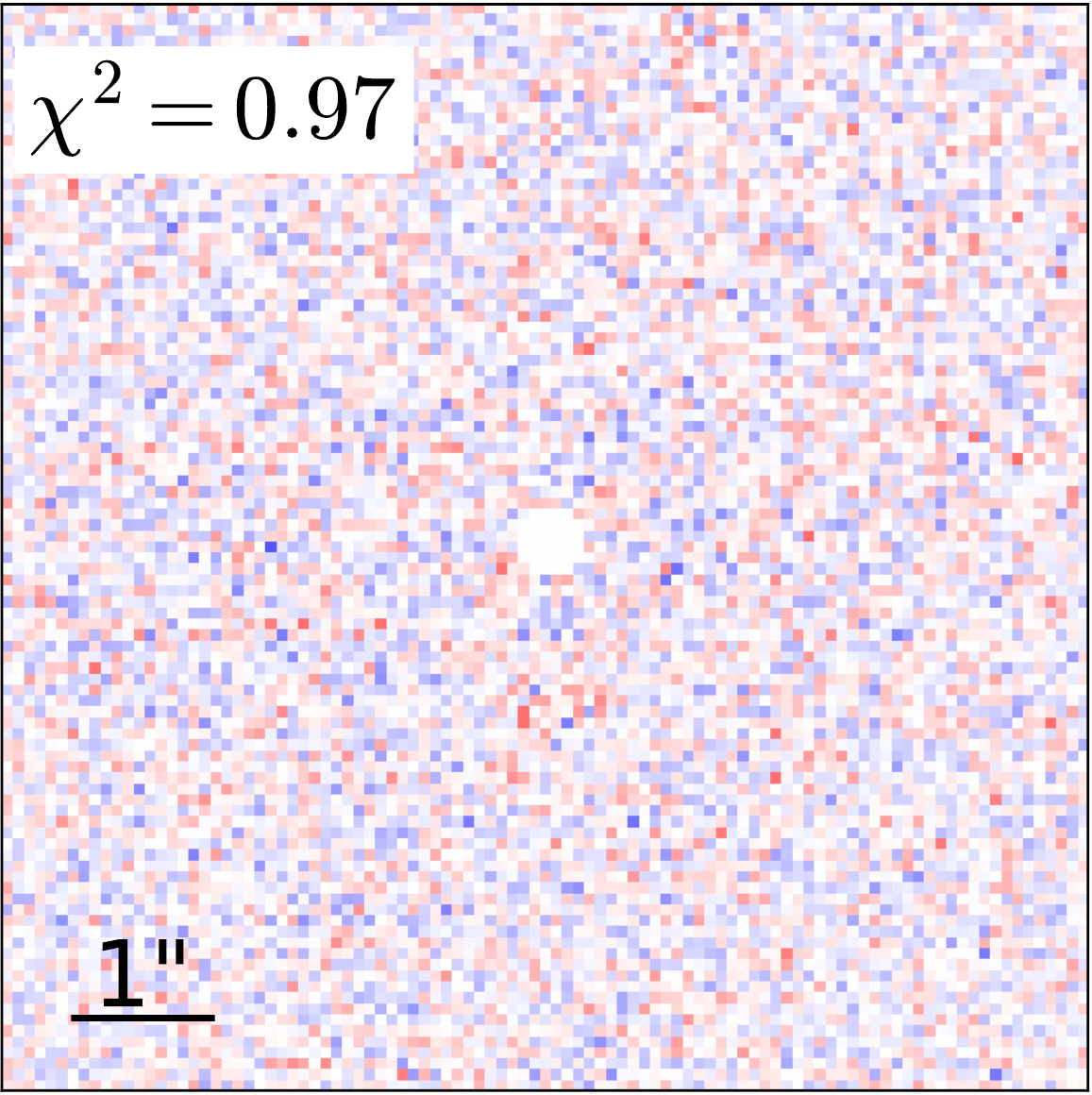} &\includegraphics[width=0.14\textwidth]{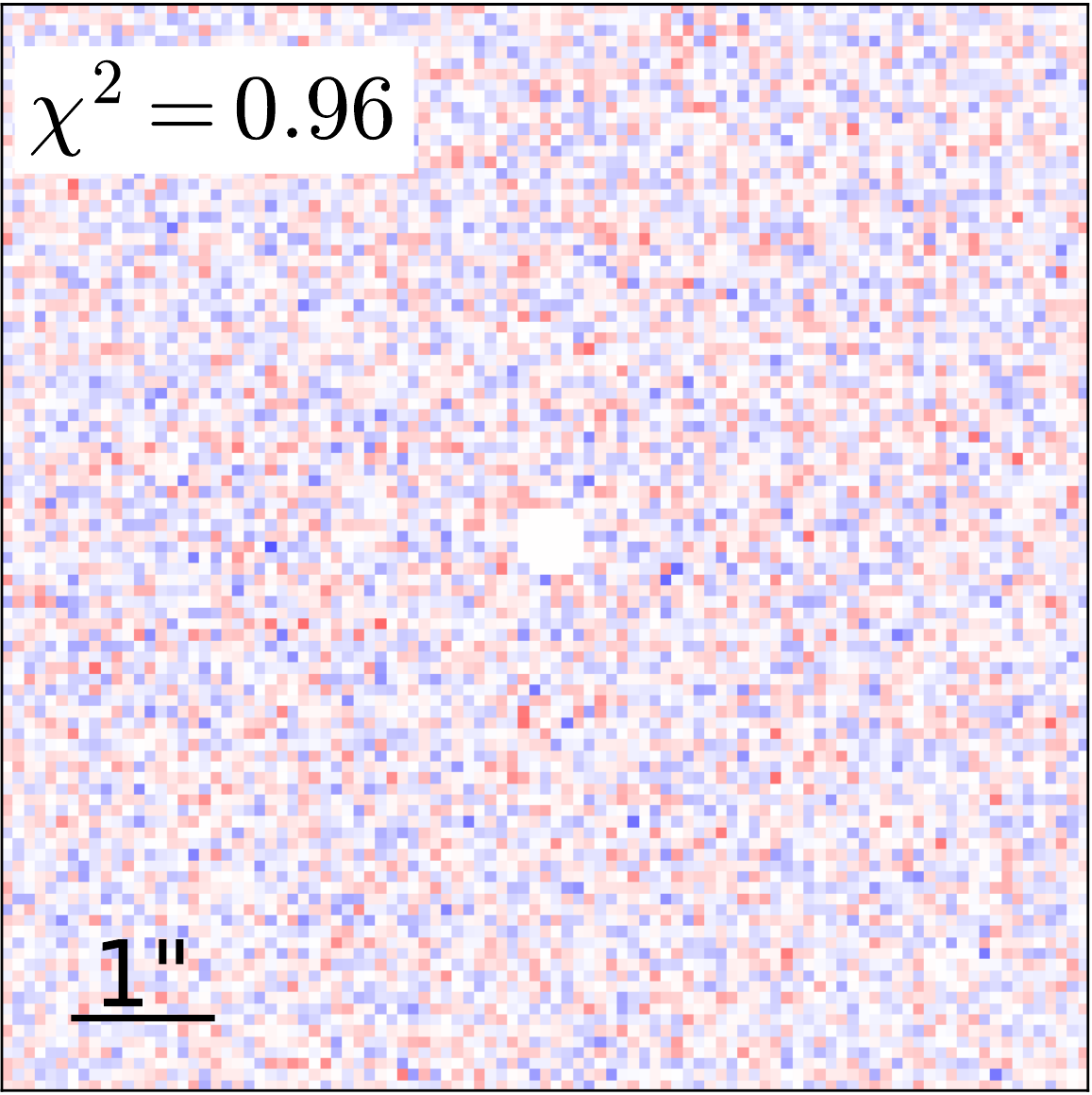} \\ 
Truth: composite (\#6) &\includegraphics[width=0.14\textwidth]{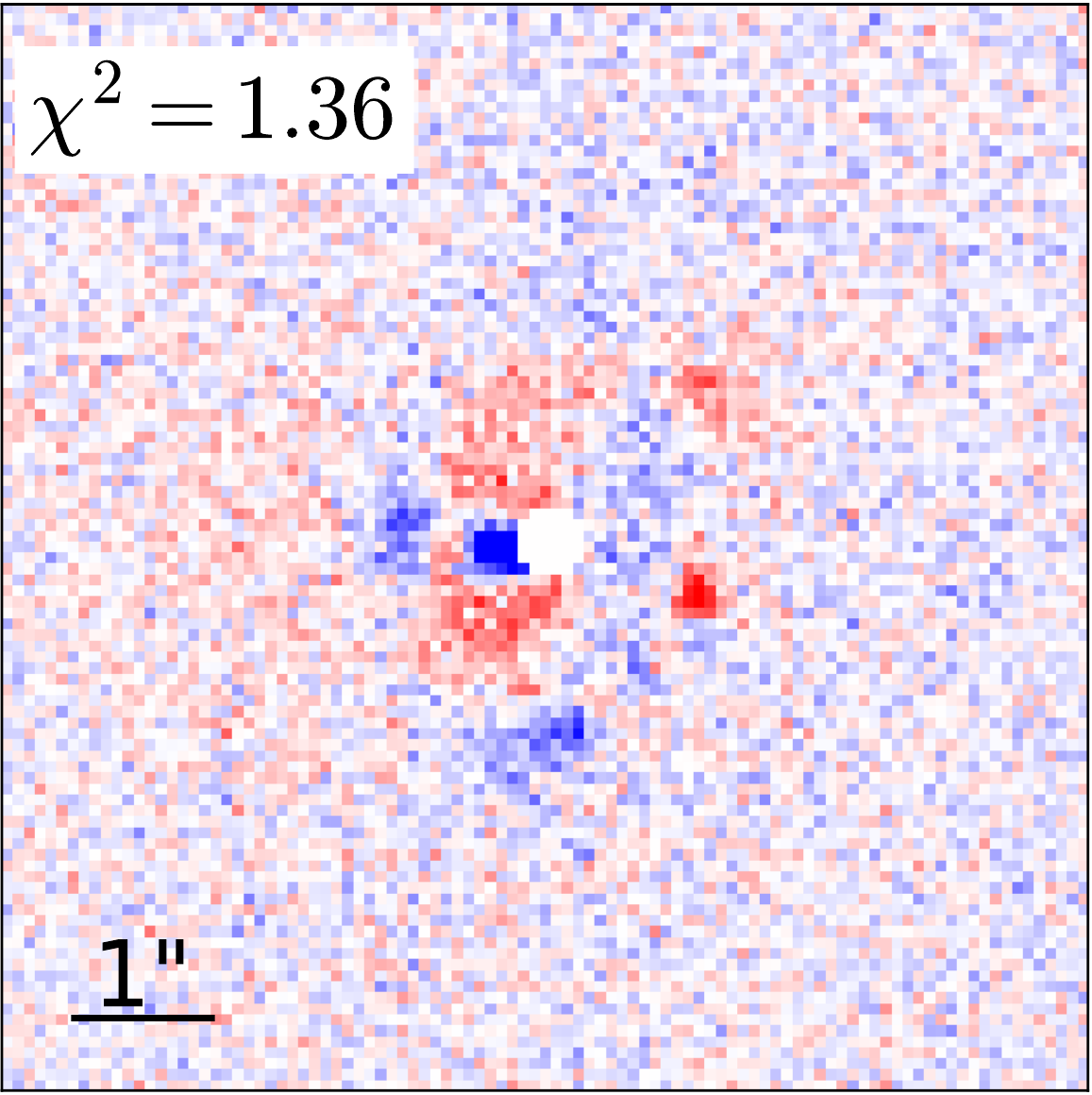} &\includegraphics[width=0.14\textwidth]{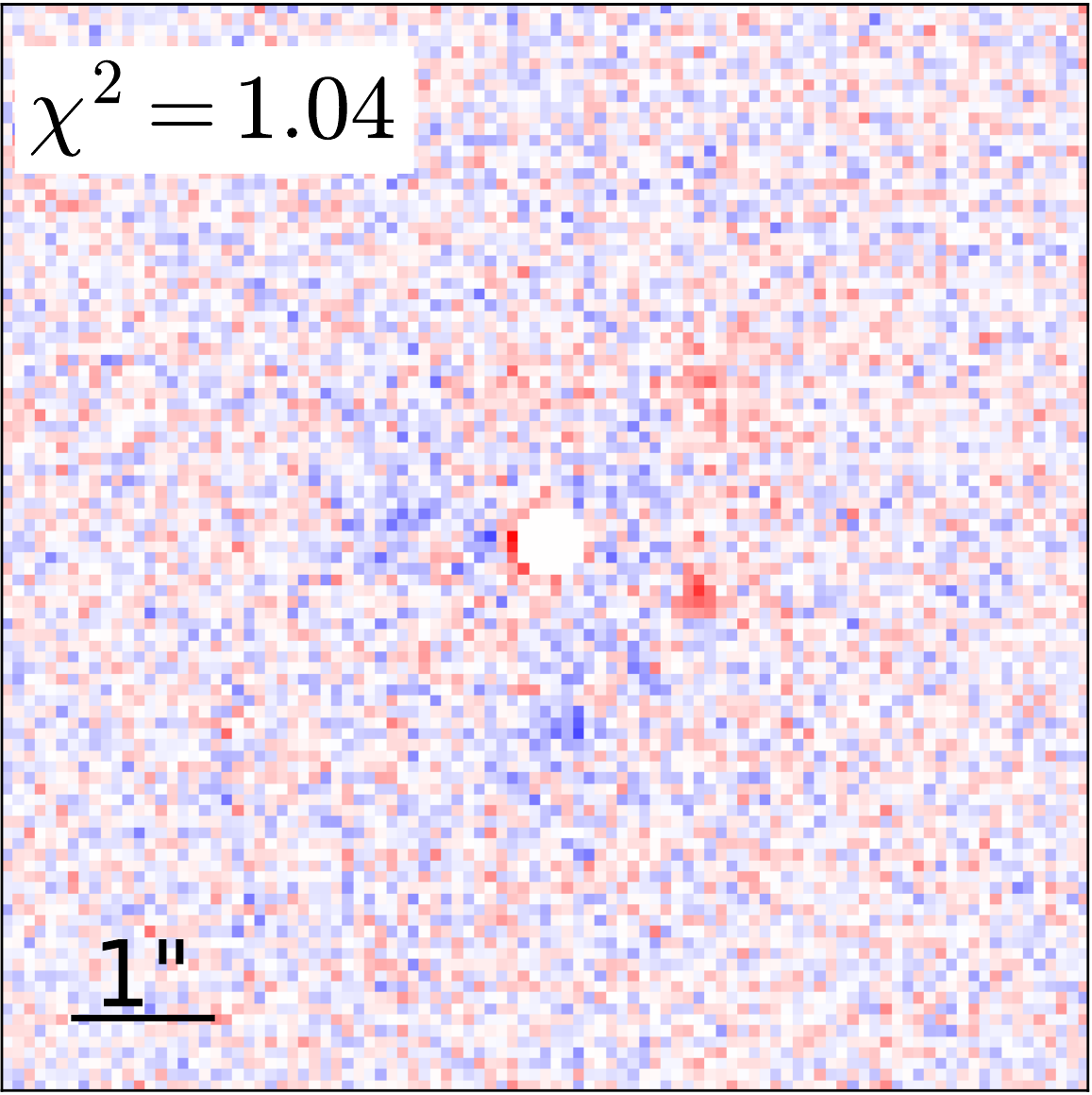} &\includegraphics[width=0.14\textwidth]{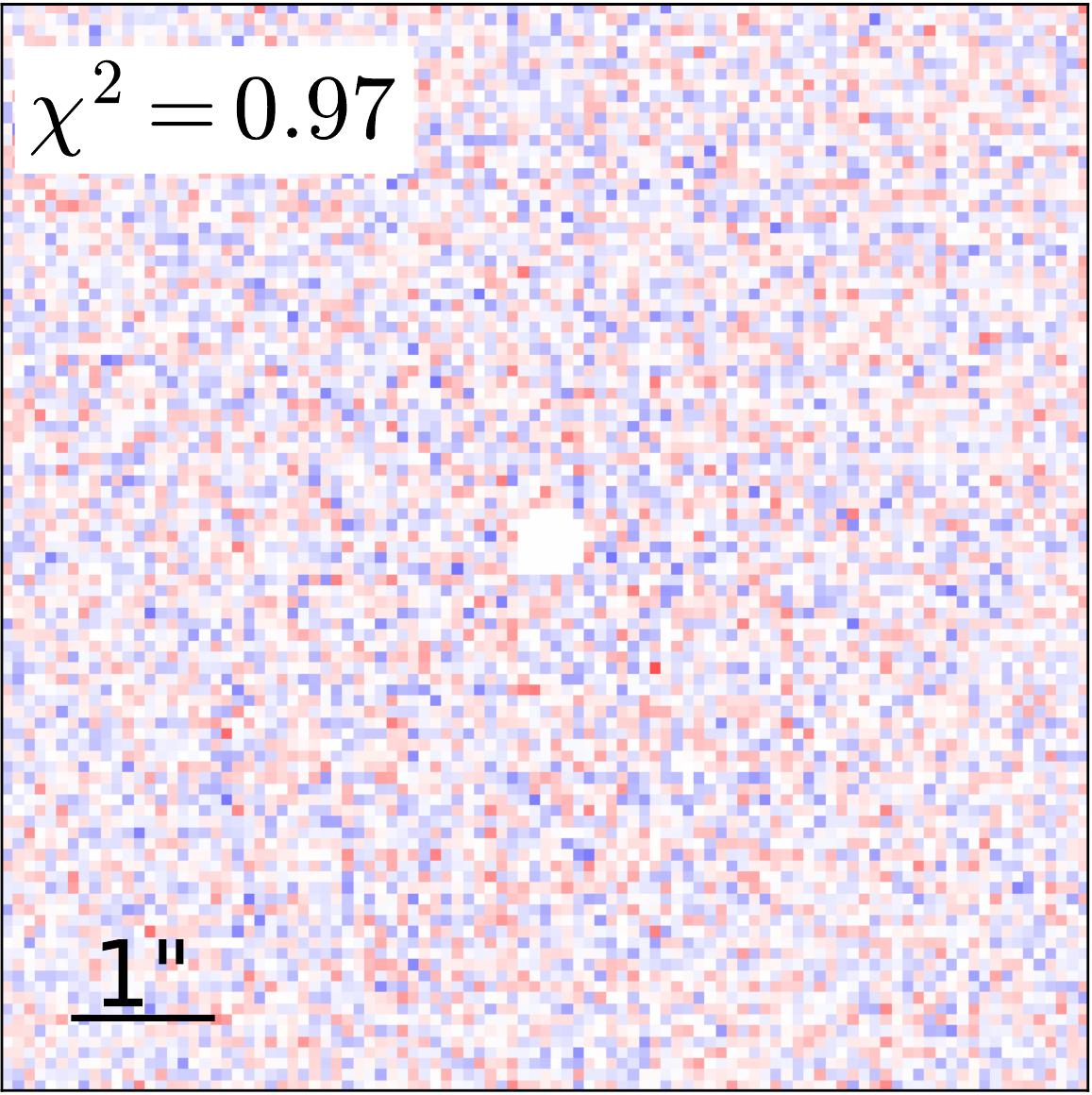} \\ 
\end{tabular}
\caption{\label{tab:test_residuals}Residual maps of the lens modelling, i.e. normalized $\chi^2$ per pixel. The maps corresponds to $(f_\mathrm{model} - f_\mathrm{data}) / \sigma$, where $f_\mathrm{data}$ is the observed flux, $f_\mathrm{model}$ is the modelled flux and $\sigma$ is the estimated $rms$ noise level at the pixel position. The color map ranges from -6$\sigma$ (blue) to +6 $\sigma$(red). The given $\chi^2$ value in each panel is the mean $\chi^2$ per pixel and does not include the time-delay information.}
\end{table*}

Interestingly, the core size of the cored power-law profile is well constrained by the data. Indeed, when a cored power-law profile is fitted to data generated with power law with no core, the core size is well constrained and shrinks to zero. If there is a core in the simulation (e.g mock lenses \#3 and \#4), the core-size is recovered within 2.2\% accuracy and within  $<$3.0\% precision with a cored power-law model. This indicates that the lensing data are sensitive to the presence of a sizeable core in galaxies.  The sensitivity stems from the robust constraint on the mass enclosed within the Einstein radius that indirectly depends on the core size.

We deliberately choose not to present the results of the composite models fitted to power-law and cored power-law simulations. This is because, by construction, the lens light profile of these simulations does not necessarily correspond to their mass profile. In the power-law and cored power-law profiles, the lens light profile bears no relation to the mass distribution, and is only used as a tracer of the stars when computing the stellar velocity dispersion. As a result, we cannot have a meaningful comparison between power-law and composite models if we assume that the baryonic component of the composite model is traced by the arbitrary lens light in the power-law model. This limitation is inherent to these simulations and we do not expect that this happens for real galaxies, because it is unlikely that the stellar light is not tracing at all the baryonic mass component.

The tests performed on composite simulated lenses \#5 \& \#6 show that the ability of a power law or a cored power law to recover the correct \Hc depends on the characteristics of the composite lenses. In both cases, the power-law models give much poorer fits to the data than the true composite models ($\Delta$BIC = 434 for \#5 and $\Delta$BIC = 4455 for \#6). Adding one more degree of freedom by using a cored power law instead of a power law improves the fit but it is still significantly poorer than the composite models ($\Delta$BIC = 95 for \#5 and $\Delta$BIC = 1049 for \#6 in the case of a cored power law). We note that the image residuals in lens \#6 are worse than that in lens \#5, since \#6 is in a fold configuration with higher lensing magnifications and thus produces correspondingly higher amounts of image residuals.
The recovered \Hc is compatible with the true value for the lens \#6, but in lens \#5 it is biased toward lower \Hc by 9.4\%. In short, the different behavior arises because of intrinsic differences in the composite mass density profile. While mock \#5 is chosen to be different from a power law, mock \#6 is chosen to be similar to a power law. When the truth is a composite similar to a power law, the inferred \Hc is the same. When it is not, the two models lead to different inferences. As discussed in~Section~\ref{ssec:powervscomposite} the real universe is similar to \#6 and dissimilar to \#5.

As an additional test, we model the simulated data using only the four lensed image positions, the lensing galaxy position and the time delays to investigate the effect of neglecting the other information. We find that, as mentioned in Section~\ref{sec:toysmodels}, this is not sufficient to constrain all the lens model parameters. A reduced $\chi^2 <$ 1 can be obtained for all the mocks using a power law model, except for mock \#6 for which the best reduced $\chi^2$ is $\sim 1.9$. Even when the true mass distribution is a power law (e.g., mock \#1 and \#2), the maximum likelihood models are associated with power-law indices substantially different from the input one, yielding a bias on \Hc , that can reach 90\% (see Appendix~\ref{app:simple_model} for details). This is well understood as the multipole components of the lens potential can compensate for large changes in the monopole structure which are only poorly constrained by the few image positions. This test highlights the necessity of using the full information provided by the high resolution images to better constrain the lens potential. In particular, the multiple images of the lensed host galaxy are critical to pin down the uncertainty on the average mass density at the image positions \citep{Kochanek2001}.

\begin{figure*}[p!]
    \centering
    \includegraphics[width=0.9\textwidth]{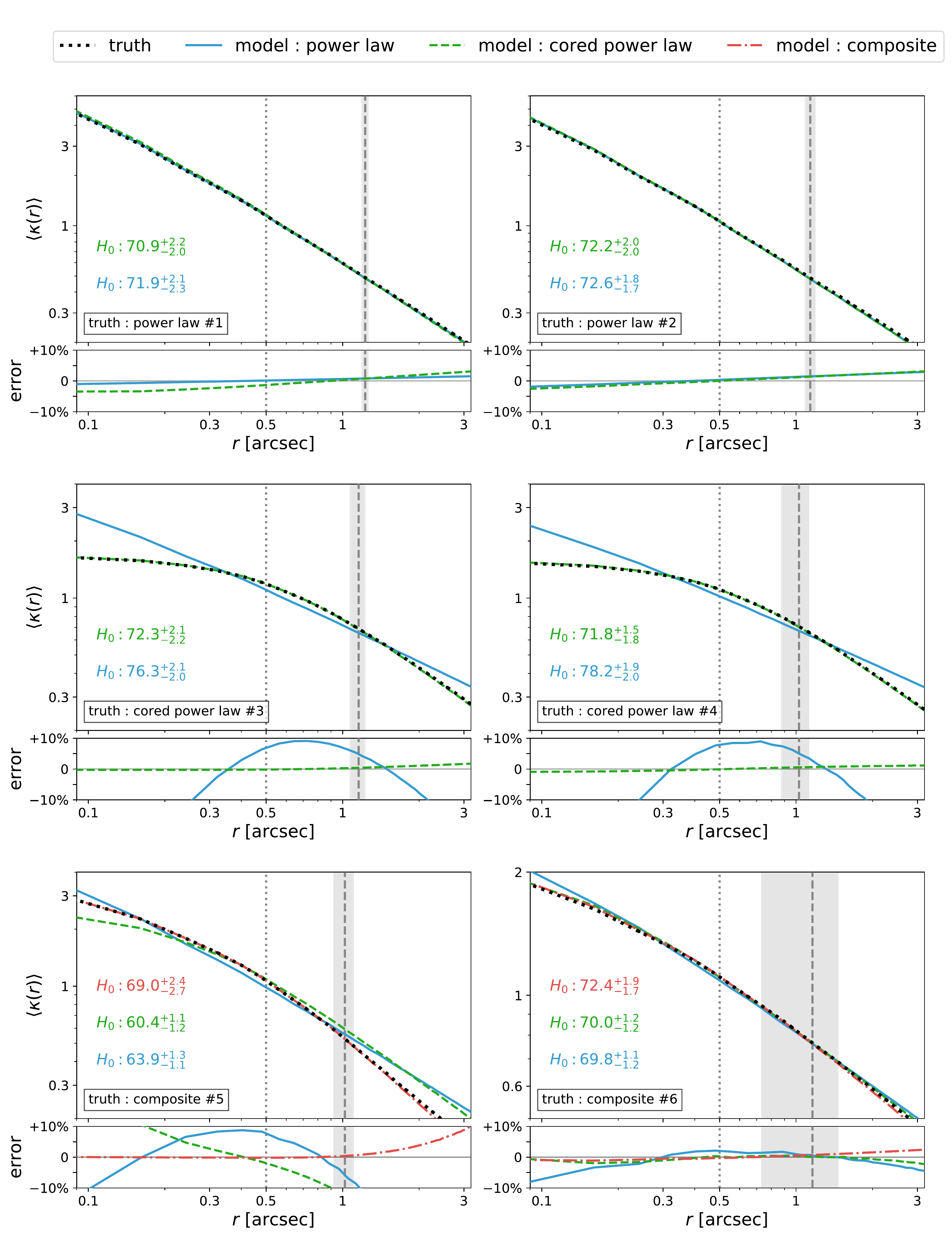}
    \caption{Azimuthally averaged radial convergence profiles, for the different lens models applied to fit the sample of six mock lenses (Fig. \ref{fig:mock_lenses}). Upper part of each panel: true profiles are shown in dotted lines; power-law, cored power-law and composite models are shown in blue continuous, green dashed and red dot-dashed lines, respectively. The spectroscopic (square) aperture used for computing velocity dispersions is indicated as a vertical dotted line, and the true Einstein radius location is indicated as a vertical dashed line. The gray area encloses lensed quasar image positions. For each model, the inferred \Hc values are indicated (in \ksmpc), and must be compared to the input value $H_0^\mathrm{fiducial}=70.7$ \ksmpc. Lower part of each panel: relative error computed as (truth $-$ model) / truth. The pixel size in the simulated images is $0\farcs08$.}
    \label{fig:kappa_profiles}
\end{figure*}

\subsection{Discussion}
\label{ssec:disc}

In this section we discuss the results of the simulations with the goal of providing an intuitive physical understanding of the quantities that are relevant for time-delay cosmology and how they are constrained by the data. As noted by \cite{Kochanek2002}, the time delay is mainly determined by the mean convergence $\langle \kappa \rangle$ in an annulus between the multiple images. Figure \ref{fig:kappa_profiles} shows the radial convergence profiles of the models averaged over the azimuth angle. The shaded gray contour corresponds to the separation between the multiple images. The quality of the fit in this region determines the accuracy on \Hc. 
The Einstein radius is typically very well constrained by any lens model, so the only way to modify the mean $\langle \kappa \rangle$ at the positions of the multiple images
is to change the slope of the convergence profile while keeping constant the integrated mass within the Einstein radius. This is a well-known problem in time-delay cosmography called the profile slope degeneracy \citep{Witt2000,Wucknitz2002,Suyu2012}.

As argued by \cite{Sonnenfeld2018}, assuming a too rigid model such as a power law model, can lead to a bias up to $\sim 10$\% if the true underlying profile contains a change of slope within the Einstein radius. \cite{Sonnenfeld2018} concluded that at least three degrees of freedom are required in the lens model to recover an un-biased result if no kinematics information is used. With the addition of kinematics, uncertainty can be reduced to 1\% (in accuracy) even within the simplified constraints considered in that study. Based on a sample of simulated galaxies from the Illustris simulation, \cite{Xu2016} studied how physically motivated numerical density profiles are transformed into an approximate power-law (in the region where lensed images are formed) by means of a mass-sheet like transformation. They reported that a large range of transformation was allowed, which would translate into a large scatter and possible bias on the inferred \Hc. They concluded that the amplitude of the bias depends on the (logarithmic) curvature of the mass density profile of the simulated galaxies. This behavior was previously illustrated in \cite{Schneider2014} and more recently in \cite{Gomer2019}.

We recover the findings of \cite{Sonnenfeld2018}, \cite{Xu2016}, \cite{Schneider2014} and \cite{Gomer2019} with our simulated lens \#5, where the combination of the Hernquist and NFW profile is designed to produce an inflection point in the radial profile of the convergence within the Einstein radius. For this system, the composite and power-law models are discrepant, thus providing an indication that the power-law model is indeed too rigid. This rigidity results in a significant difference in goodness of fit ($\Delta$BIC=434), as well as on the inferred \Hc.

For the lens system \#6, the radial convergence profile does not have inflection points and therefore it is impossible to change the slope of the profile while keeping the Einstein radius identical. In this case, the recovered value of \Hc is compatible with the true value for both the composite and power-law model. The fact that the two families of models are providing compatible \Hc indicates that the convergence profile is well-recovered in the annulus around the Einstein radius. 

The TDCOSMO collaboration has systematically tested both model families in their analysis after the first and only nonblind published system \Bsixteen.
The tight agreement between the composite and power-law models in the TDCOSMO analyses supports the hypothesis that, as a result of the bulge-halo conspiracy, the kind of real galaxies that act as strong lenses are similar to our \#6 mock. The mass density profile is well approximated by a power law. In this case, the stellar component and the extended NFW halo combine to form a profile very close to a power law near the quasar images. If this had not been the case, we have shown in this work that the composite and power law would not have produced the same mean convergence $\langle \kappa \rangle$ at the image position, and thus would have yielded very different \Hc values. If TDCOSMO had found this discrepancy, it would have been accounted for in the error budget of each individual lens since they marginalize over model families. In contrast, both classes of models produce fits with comparable goodness-of-fit and \Hc in the real world, resulting in high precision, including modeling errors in the error budget. Our analysis in Section~\ref{ssec:powervscomposite} shows that in practice the two models agree even at the current sample precision of $<$2.4\%.

%%%%%%%%%%%%%%%%%
\section{Conclusion}
\label{conclusion}
%%%%%%%%%%%%%%%%%

As the statistical precision of time-delay cosmology improves with the analysis and publication of multiple gravitational lens systems by the H0LiCOW, SHARP, and STRIDES collaborations, a parallel effort must be undertaken to ensure that systematic uncertainties remain subdominant. In this first paper of the TDCOSMO collaboration (i.e., COSMOGRAIL, H0LiCOW, SHARP, STRIDES members), we investigate and quantify a number of potential systematic uncertainties that could affect the analysis. Before we summarize the main results of this work, it is important to highlight a few general points that are relevant to the estimation of systematic errors in time-delay cosmography:
\begin{enumerate}
    \item The TDCOSMO analyses are carried out blindly to cosmological parameters, with the exception of the first system \Bsixteen \citep{Suyu2010} in order to avoid implicit/explicit experimenter bias.
    \item The TDCOSMO estimates of \Hc are obtained independently for each lens, and they are found to be statistically consistent with each other \citep{Wong2019}. The statistical consistency demonstrates that uncorrelated systematic errors are negligible with respect to statistical errors. So any investigation of systematic errors must focus on correlated errors that would affect many systems in the same way.
    \item Toy models based on simplified assumptions and constraints cannot offer any quantitative estimates of systematic errors given the current state-of-the-art data-sets and lens models. The only way to estimate quantitative errors is to carry out an analysis that is very similar to the one performed on real data, using the full extent of the available information, including the high-resolution images, multiple time delays (if available), and stellar kinematics. For example, the dependency on the inferred distances on stellar velocity dispersion is nontrivial, it varies from lens to lens, depending on the precision of the various constraints, the lensing configuration, the source and deflector redshift, and the spectroscopic aperture used for the kinematic measurement. On average over the current TDCOSMO sample, uncertainty in velocity dispersion $\delta \sigma_v / \sigma_v$ translates into approximately $\delta H_0 /H_0 \sim 0.07 \times \delta \sigma_v / \sigma_v$. %\shscom{edited last sentence}
\end{enumerate}

Keeping these general considerations in mind, the main results of this work are as follows:

\begin{itemize}
    \item No evidence is found for any correlation between the measured value of \Hc and observables related to the internal structure of the lens galaxies (e.g., velocity dispersion, effective radius), or to the size of the spectroscopic aperture. If our assumptions about the kinematic field of the lens galaxies had been significantly wrong, then we would have expected to detect trends in these parameters, since our deflectors and spectroscopic observations span a significant range of configurations. Of course absence of evidence is not evidence of absence and more work remains to be done in this area, even though the weak dependency of the inferred \Hc on kinematic data implies that systematic uncertainties in this area will have a subdominant impact on \Hc.
    
    \item No evidence is found for any correlation between the measured value of \Hc and the external convergence estimated from galaxy number counts and numerical simulations. In contrast, if no external convergence is applied, \Hc is found to depend on the overdensity of galaxies in the field, a clearly unphysical result.

    \item Tests based on mock lens systems that have simulated data comparable in quality to real lens systems show that the current approach of considering different mass profiles has sufficient flexibility in the mass model to infer a wide range of \Hc values, should the data require it.
    
    \item Mock lens galaxies composed of baryons and dark matter whose total mass distribution is not well approximated by a power law produce discrepant \Hc inferences and significant differences in image residuals when comparing power-law and composite mass models. In contrast, mock lens galaxies whose baryonic and dark matter components conspire to form a power law lead to comparable \Hc inferences between power-law and composite mass models.
    
    \item The comparison of power-law and composite mass models allows us to quantify deviations in \Hc due to our mass model assumptions. By using these two families of models and marginalising over them, the resulting \Hc accounts for modeling uncertainties. Future measurements of spatially resolved kinematics of the lens system would provide highly constraining measurements of the lens mass distribution that potentially allow us to distinguish/rank mass models, removing the need to marginalize over degenerate lensing mass models.
    
    \item The similarity of \Hc constraints from power-law and composite models of TDCOSMO lenses shows that the total mass profiles of galaxies are close to power laws due to the bulge-halo conspiracy.  For the six lenses that have been analyzed with both power-law and composite models we find \Hc=$74.2^{+1.6}_{-1.6}$ \ksmpc and $74.0^{+1.7}_{-1.8}$ \ksmpc respectively. The difference between the two model families is much smaller than the inferred statistical errors. The similar \Hc from the different families of models thus made the current \Hc measurement with $\sim$2\% uncertainty from TDCOSMO achievable.
    
\end{itemize}

Based on a number of tests carried out in this paper, we find no evidence that the error budget reported by the H0LiCOW/SHARP/STRIDES (TDCOSMO) collaborations is significantly underestimated. We emphasize that our tests reproduce very closely the TDCOSMO inference procedure, in contrast to previous work in the literature that does not have the fidelity to investigate this issue.

While investigating potential sources of systematic uncertainties was an important first step, meeting the goal of 1\% precision and accuracy with time-delay cosmography \citep[e.g.,][]{Suyu2012,TreuMarshall2016}, requires additional and continued efforts over the coming years. Aside from expanding sample sizes and improving statistical precision per system, some of the clear steps along the way are:  i) exploring broader model families and the impact of departures from elliptical symmetry and including spatially variable mass-to-light ratio, ii) explore in more detail the bulge-halo conspiracy based on high resolution data for local early-type galaxies, iii) explore the effect of allowing for gradients in stellar mass-to-light ratios \citep[e.g.,][]{Sonnenfeld+18} in composite models; iv) carrying out a full Bayesian hierarchical analysis of existing samples of lenses in order to constrain parameters that cannot be inferred on single lens but require an inference at the population level, v) accounting for measurement and modeling covariance, and vi) performing realistic data challenges such as the one proposed by \citet{Ding2018}, with increasing level of realism and complexity as data also improve. These steps are nontrivial from a modeling point of view, considering that the analysis of any single system currently requires a year of expert investigator time and on the order of a million CPU hours \citep[e.g.,][]{Shajib2019}. Substantial advances in automation and speed are required in order to carry out those next steps, but given their importance for the determination of \Hc, they are worth undertaking. 

\begin{acknowledgements}
This program is supported by the Swiss National Science Foundation (SNSF) and by the European Research Council (ERC) under the European Union’s Horizon 2020 research and innovation program (COSMICLENS: grant agreement No 787886). Additional funding is provided by the Packard Foundation through a Packard Research Fellowship to T.T., by the National Science Foundation through grant AST-1906976 to T.T.~and by NASA through HST grants HST-GO-10158, HST-GO-12889, HST-GO-14254, HST-GO-15320, HST-GO-15652.  S.H.S.~thanks the Max Planck Society for support through the Max Planck Research Group. G.C.-F.C. acknowledges support from the Ministry of Education in Taiwan via Government Scholarship to Study Abroad (GSSA). C.D.F. and G.C.-F.C.~ acknowledge support for this work from the National Science Foundation under Grant Nos. AST-1715611 and AST-1907396.  This work was supported by World Premier International Research Center Initiative (WPI Initiative), MEXT, Japan. L.V.E.K.~has been supported through an NWO-VICI grant (project number 639.043.308). This research made use of Astropy, a community-developed core Python package for Astronomy \citep{Astropy2013, Astropy2018}, the 2D graphics environment Matplotlib \citep{Hunter2007} and {\tt emcee}, a Python implementation of an affine invariant MCMC ensemble sampler \cite{Foreman2013}. 

\end{acknowledgements}

\bibliographystyle{aa}
\bibliography{biblio}

\begin{appendix}
\onecolumn
\section{Properties of simulated lenses}
As a complement to Section~\ref{sec:6}, we show in Table~\ref{app:mock_params} a subset of important properties of the simulated lenses. Characteristic radii are indicated: half-light radius, effective Einstein radius, core radius in the case of cored power-law profiles, and scale radius of the dark matter profile for composite models. The ratio of the lens half-light radius and Einstein radius is also computed. Additionally, the input logarithmic slope of the convergence profile, the lens mass ellipticity, true time delays and LOS velocity dispersion are indicated. The spectroscopic aperture used for simulating and modeling kinematics is a square with side $1''$. For composite models, we provide the dark matter fraction within the Einstein radius. Lastly, to ease comparison with previous studies, we add the measure of the curvature of the total mass $\xi$, as defined in \cite{Xu2016}. Given this definition, a concave-upward (convex-downward) radial convergence profile has curvature greater than one (lower than one), and a perfect power-law have curvature equal to one. 
\cite{Xu2016} conclude that galaxies close to isothermal and those having a curvature parameter close to one provide the smallest bias on \Hc. We recover these findings only partially with these simulated lens systems. We find that the curvature criterion is the most important criterion to ensure a low bias on \Hc even if the slope differs significantly from isothermal, as illustrated with our simulated galaxy \#6.

\begin{table*}[htbp!]
\renewcommand{\arraystretch}{1.2}
\centering
{\small
\begin{tabular}{l|c|c|c|c|c|c|c|c|c|c|c}
{} &  $\theta_{\rm eff}$\ [$\arcsec$] &  $\theta_{\rm E}$\ [$\arcsec$] &  $\theta_{\rm E} / \theta_{\rm eff}$\  &  $\theta_{\rm c}$\ [$\arcsec$] &  $r_{s}$\ [$\arcsec$] &  $\gamma$\  &      $q$\  & $\xi$\  &  $f_{\rm DM}$ &      $\Delta t$\ [days] &  $\sigma_v$\ [km\,s$^{-1}$] \\
\hline\hline
\#1 power law       &                     3.620 &                   1.237 &                                  0.342 &                       - &              - &                  2 &      0.899 &   1.000   & -    &     [0.277, 3.701, 8.999] &            308 \\
\#2 power law       &                     3.789 &                   1.143 &                                  0.302 &                       - &              - &                  2 &      0.889 &    1.000 & -     &   [3.919, 4.48, 10.773] &              297 \\
\#3 cored power law &                     3.988 &                   1.153 &                                  0.289 &                   0.559 &              - &                  2 &      0.890 &    1.016 & -    &    [1.331, 5.687, 7.112] &              245\\
\#4 cored power law &                     2.643 &                   1.028 &                                  0.389 &                   0.560 &              - &                  2 &      0.895 &    1.018  & -   &   [3.135, 3.525, 9.012] &              216 \\
\#5 composite       &                     2.689 &                   1.028 &                                  0.382 &                       - &         31.185 &                    1.90 &      0.900 &     1.036  & 0.190  &   [3.55, 9.175, 13.567] &               253 \\
\#6 composite       &                     3.573 &                   1.165 &                                  0.326 &                       - &         34.497 &                    1.45 &      0.902 &   1.006  & 0.763   &   [4.878, 5.055, 12.166] &              207 \\
\end{tabular}
}
\caption{Key properties of the simulated lenses described in Section~\ref{sec:6}. For each lens, from left to right: lens half-light radius $\theta_{\rm eff}$, effective Einstein radius $\theta_{\rm E}$ (enclosing a mean convergence equal to unity), ratio of these radii, core radius $\theta_{\rm c}$, dark matter scale radius $r_{s}$, effective slope of the convergence profile at the Einstein radius $\gamma$, lens mass ellipticity $q$, total mass curvature $\xi$ \citep[as defined in][]{Xu2016}, dark matter fraction $f_{\rm DM}$ within Einstein radius, true time delays $\Delta t$ and LOS velocity dispersion $\sigma_v$ of the lens galaxy.}
\label{app:mock_params}
\end{table*}

\section{Models only based on lensed quasar positions and time delays}\label{app:simple_model}

We model the simulated dataset as generated in Section~\ref{ssec:sim}, using only the lensed quasar positions, the lensing galaxy position and the relative time delays between the lensed images. We assume an uncertainty $\sigma = 0\farcs{004}$ on the point-source positions, $\sigma = 0\farcs{01}$ on the lensing galaxy centroid, and the same uncertainty on the time-delay as in Section~\ref{ssec:results}. Similarly to extended source modeling performed in Section~\ref{ssec:results}, we employ the lens modeling package \LENSTRO. We adopt both the Singular Isothermal Ellipsoid (SIE) model (i.e., fix slope value $\gamma=2.0$) and the Power-law model in this test. An independent modeling has been carried out with \texttt{lensmodel} \citep{Keeton2001, Keeton2011}. We obtained similar inference with both packages and therefore only report hereafter results obtained with \LENSTRO.  

We use the true parameters as the input values to start performing the modeling. A careful choice of the likelihood and sampling options has to be carried out to ensure that image position constraints arise from the same source. In practice, we sample the source plane and evaluate the positional likelihood in the image plane, but adding a source plane likelihood term to ensure that each image arises from the same source within $\sigma = 0\farcs{001}$. A notebook implementing our fitting strategy is available at \url{https://github.com/TDCOSMO/TD_data_public/blob/master/TDCOSMO_I/PSTD_notebook.ipynb}.

\begin{figure*}[t!]
    \centering
    \includegraphics[width=0.49\textwidth]{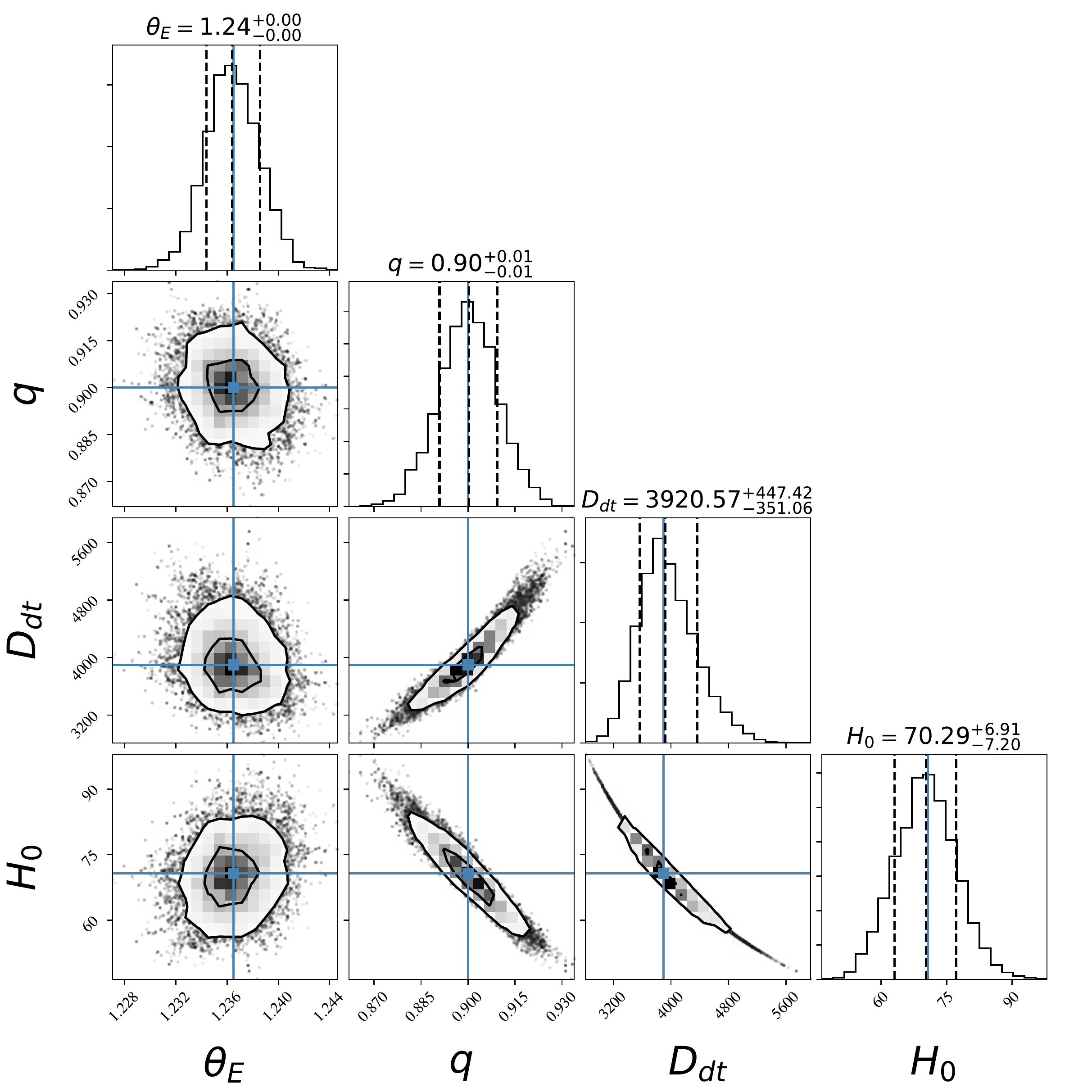}
     \includegraphics[width=0.49\textwidth]{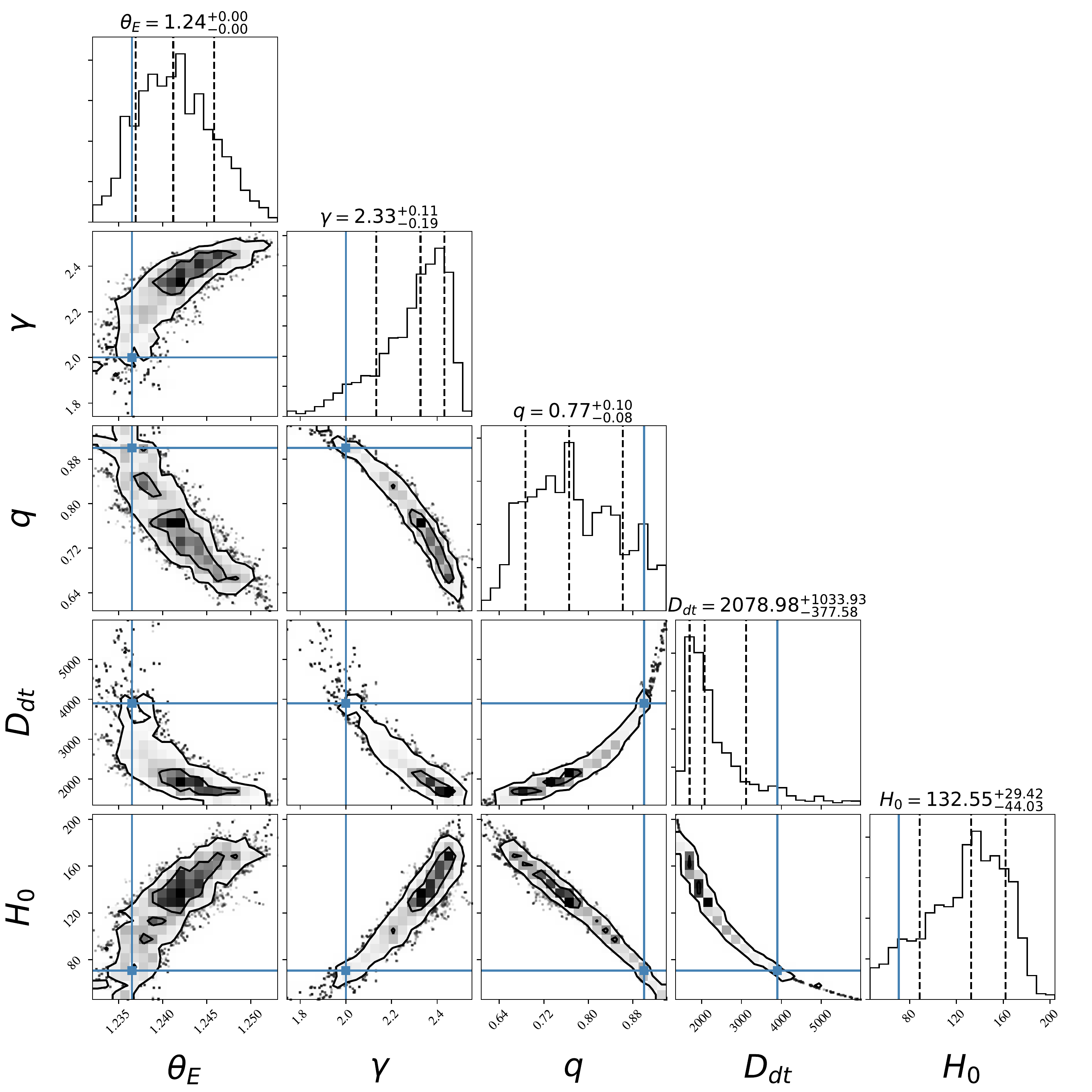}
    \caption{The corner plot of the inference of modeling mock system \#1 using only the lensed quasar position and time delay. The SIE model and Power-law model are adopted on the left and right, separately. The blue lines indicate the true values in the simulation.}
    \label{fig:corner_pos_only}
\end{figure*}

In Fig.~\ref{fig:corner_pos_only}, we show the corner plots of the inference based on the mock system \#1. When using a SIE model where the mass slope value is fixed to the truth, we could obtain an unbiased \Hc\ with uncertainty at the $\sim10\%$ level. However, when the slope is a free parameter, the inference broadens significantly as the data are not sufficient to constrain that parameter. In particular, the uncertainty on \Hc increases by a factor of $\sim3$ and the maximum likelihood deviates from the truth by up to 100\%. This contrasts with the same model constrained by the point-source and extended images from the source. Those features constrain accurately the position of the centroid of the lens potential and its ellipticity, breaking degeneracies between those quantities and $H_0$. 

Qualitatively similar behavior is observed for the other systems, but we do not report inferred parameters in those cases due to the difficulty to achieve convergence of the MCMC chains for the power-law model. This is due to the degeneracies observed between $q$, $\gamma$ and $H_0$ which implies a sampling of a large region of the parameter space, enforcing exploration of parameter values for which results of \LENSTRO \, modeling has not been fully tested (e.g., $\gamma > 2.5$). 

\end{appendix}
\end{document}